# Functional Optical Coherence Tomography for Intrinsic Signal Optoretinography: Recent Developments and Deployment Challenges

Tae-Hoon Kim[1], Guangying Ma[1], Taeyoon Son[1], and Xincheng Yao[1,2*]

[1]Department of Biomedical Engineering, University of Illinois at Chicago, Chicago, Illinois, United States

[2]Department of Ophthalmology and Visual Sciences, University of Illinois at Chicago, Chicago, Illinois, United States

***Correspondence:** Xincheng Yao (xcy@uic.edu)

## Abstract

Intrinsic optical signal (IOS) imaging of the retina, also termed as optoretinography (ORG), promises a noninvasive method for objective assessment of retinal function. By providing unparalleled capability to differentiate individual layers of the retina, functional optical coherence tomography (OCT) has been actively investigated for intrinsic signal ORG measurements. However, clinical deployment of functional OCT for quantitative ORG is still challenging due to the lack of a standardized imaging protocol and the complication of IOS sources and mechanisms. This article aims to summarize recent developments of functional OCT for ORG measurement, OCT intensity- and phase-based IOS processing. Technical challenges and perspectives of quantitative IOS analysis and ORG interpretations are discussed.

## 1 Introduction

The retina is a neurovascular network complex which can be frequently affected by eye diseases such as age-related macular degeneration (AMD), diabetic retinopathy (DR), glaucoma, and inherited retinal dystrophies (IRDs). Optical imaging methods, such as light fundus photography and fluorescein angiography [1; 2; 3], can reveal morphological abnormalities for eye disease diagnosis and treatment assessment. Scanning laser ophthalmoscopy (SLO) [4; 5] can provide improved spatial resolution and image contrast. Optical coherence tomography (OCT) [6; 7] and OCT angiography (OCTA) [8; 9; 10; 11] can provide depth-resolved observation of retinal neurovasculature. Adaptive optics can be incorporated to enhance the resolution of the fundus camera [12; 13], SLO [14; 15], and OCT [16; 17; 18]. These methods for morphological imaging of the retina provide indispensable information for clinical management of eye diseases.

However, the retinal diseases are often quite advanced before they draw clinical attention by which time the photoreceptors may be functionally abnormal. Structural and functional abnormalities in the retina are not always correlated in terms of the time window and spatial location. Therefore, objective functional assessment of the retina promises early detection and longitudinal therapeutic assessment of retinal degenerative diseases. Electroretinography (ERG) and multifocal ERG [19; 20] can provide objective assessment of retinal physiological function. However, separate morphological imaging and functional measurement can be costly and time-consuming. Moreover, different spatial resolutions of morphological imaging and functional measurement may challenge the clinical evaluation.

Intrinsic optical signal (IOS) imaging of the retina [21; 22; 23; 24; 25; 26; 27; 28; 29], also termed as optophysiology [30], optoretinogram [31; 32; 33; 34; 35] or optoretinography [25; 36; 37; 38; 39; 40] (ORG), promises a noninvasive method for objective assessment of retinal physiological function. The terminology ORG is analogy to ERG. ERG is based on electrical measurement of stimulus-evoked electrophysiological activities, while ORG refers to IOS imaging of corresponding light property changes in the retina due to functional activity. Time-lapse light microscopy and fundus camera have been used for two-dimensional IOS imaging study of isolated retinal tissues and intact eyes [21; 22; 28]. By providing unparalleled capability to differentiate individual layers of the retina, OCT has been actively used for IOS imaging of animal and human retinas [25; 29; 31; 32; 33; 34; 35; 36; 37; 38; 39; 40; 41; 42; 43; 44; 45]. However, clinical deployment of the OCT based ORG is still challenging due to the lack of a standardized imaging protocol and the complication of signal sources and physiological mechanisms. In following sections, we will summarize recent developments of functional OCT system for IOS imaging, OCT intensity and phase-based processing for quantitative IOS analysis. Technical challenges and perspectives of quantitative ORG measurement and interpretation will be discussed.

## 2 Functional OCT developments for intrinsic signal ORG

The retina is thin, mostly transparent and stratified into distinct cellular layers. Since stimulus-evoked retinal activity was found to alter intrinsic optical properties at different layers, OCT has been actively investigated for IOS imaging [22; 34]. Both time-domain and Fourier-domain OCT systems have been demonstrated for IOS imaging. Time-domain OCT was first demonstrated for depth-resolved observation of IOS changes in freshly isolated frog [29] and rabbit [30] retinas. Fourier-domain OCT was later employed to validate *in vivo* IOS imaging of intact rat eyes [46]. A hybrid line-scan SLO and OCT system was also employed for confocal-OCT IOS imaging study of isolated retinas [47]. Given the improved imaging speed, Fourier-domain OCT has dominated recent IOS imaging study of both animal [33; 35; 37; 39; 42; 44; 48; 49; 50; 51; 52] and human [25; 31; 32; 36; 37; 53; 54; 55] retinas.

**Figure 1** shows representative Fourier-domain OCT system used for IOS imaging of the human retina. The system is point-scan spectral domain OCT, which is commonly used for both clinical and research purpose. The fixation target is used to minimize the voluntary eye movements during OCT imaging. The system consists of two light sources, one near infrared superluminescent diode (SLD) for OCT imaging, and a visible light source to produce retinal stimulation. The OCT probe beam is raster scanned across the tissue to image the retinal cross-section. For aiding retinal localization, a pupil camera is used, and to verify the time course of pupillary response, corresponding to the retinal stimulation. The recording speed was 100 B-scan/s with 500 A-lines/B-Scan at 70kHz A-scan rate. Point-scan OCT benefits from the confocal aperture of the single-mode fiber that rejects multiply scattered light. However, imaging speed is limited for volumetric data acquisition.

Line-scan [40] or full-field [56] parallel OCT can significantly improve the imaging speed by parallel acquisition of lateral and axial information. Full-field OCT allows imaging without lateral phase noise by employing a collimated illumination over the retinal area with detection by a 2D camera. High-speed 3D imaging can reduce intraframe eye movement artifacts, which permits robust registration of frames and tracking of photoreceptors, returning stable OCT phase information. Access to stable phase information allows detection of cellular deformations much smaller than its axial resolution, and the phase information has been recently used to measure light-evoked photoreceptor outer segment (OS) deformation. However, the parallel OCT suffers resolution loss from multiple scattering crosstalk. In addition, a tradeoff for increasing imaging speed is a reduction in imaging area.

Adaptive optics (AO) can be incorporated to further improve the OCT spatial resolution [36; 40]. The AO-OCT system consists of Fourier-domain OCT and an AO subsystem, which incorporated in the sample arm. The AO subsystem generally consists of three elements including the wavefront sensor, the deformable mirror, and the control computer to dynamically measure and correct low- and high-order wavefront aberrations of the eye. The current state-of-art AO-OCT system has resolution to reveal the 3D reflectance profile of individual cone photoreceptor and provide sufficient sensitivity to detect light-evoked optical path length (OPL) changes as small as 5 nm in the individual cones [36]. Azimipour et al. further demonstrated a combined AO-SLO-OCT for ORG measurement of rod and cone photoreceptors in the human retina [31]. The SLO was used to guide the location and type of photoreceptors in the OCT volume. However, clinical deployment of AO-OCT is still challenging due to technical factors such as high-cost, optical complexity, system size, data volume, and image postprocessing [57].

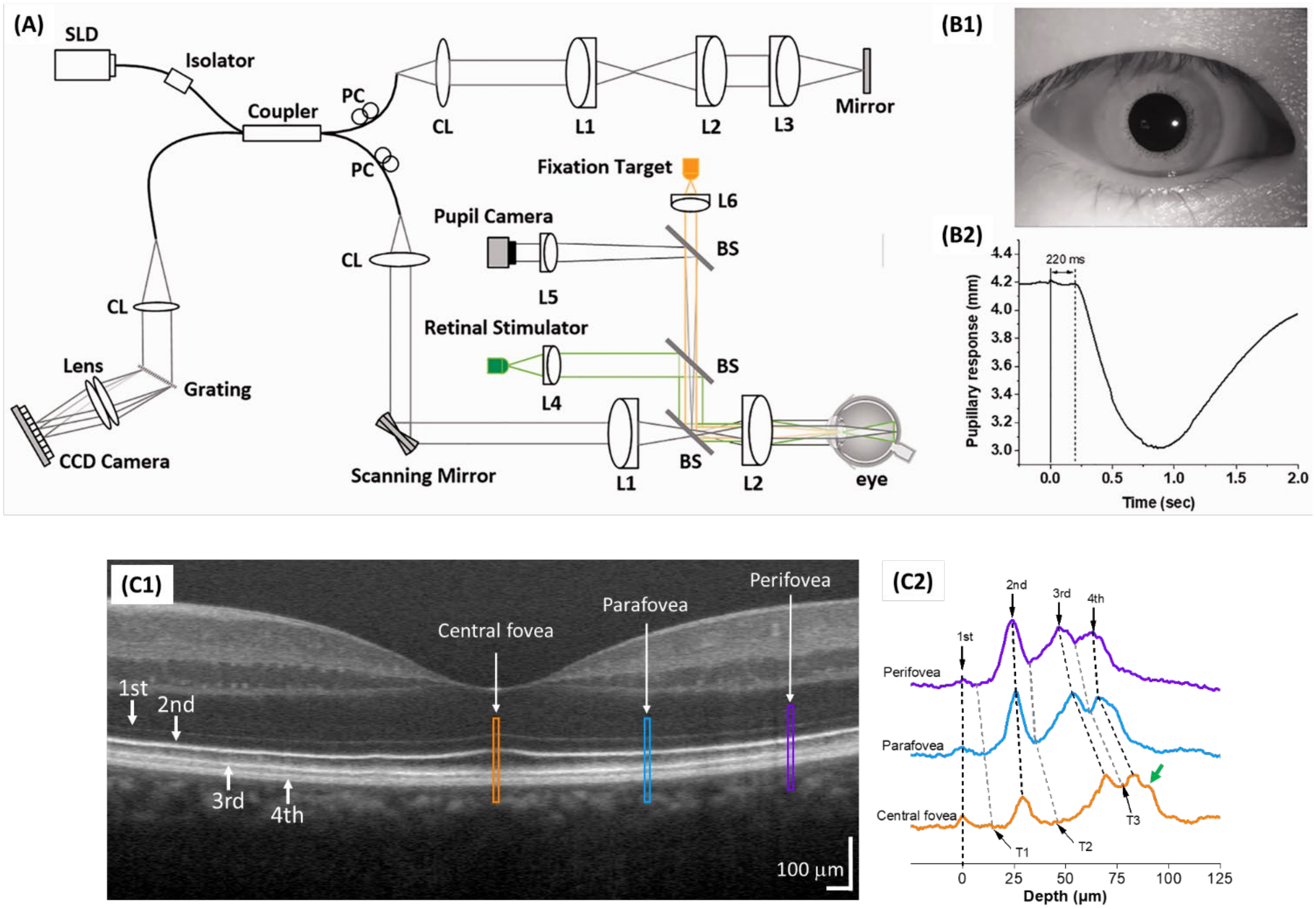

**Figure 1.** (A) Representative functional OCT system for IOS/ORG imaging. BS: beam splitter; CL: collimation lens; Lenses: L1, L2, L3, L4, L5, and L6; PC: polarization controller; SLD: superluminescent diode. (B) Representative pupil image (B1) and pupillary response (B2). Reprinted with permission from Son et al [25]. (C) Representative OCT image of human retina. NFL: nerve fiber layer; GCL: ganglion cell layer; IPL: inner plexiform layer; INL: inner nuclear layer; OPL: outer plexiform layer; ONL: outer nuclear layer; ELM: external limiting membrane; ISe: inner segment ellipsoid; OS: outer

segment, IZ: interdigitation zone, RPE/BrM: retinal pigment epithelium/Bruch's membrane. From Xincheng Yao's lab image gallery.

## 3 OCT data processing for intrinsic signal ORG

OCT is an interferometric imaging technique that acquires interference fringe patterns generated by superposition of back-scattered lights from the sample and reference arm. The Fourier transform of the fringe patterns provide not only intensity information on the scattering object but also allows access to information about the axial position of the scattering object within the retina. Both intensity and phase-based processing methods have been developed to quantify the stimulus-evoked IOS changes in the retina.

### 3.1 OCT intensity-based IOS processing

The OCT intensity-based IOS processing can be divided into two categories: OCT brightness change and OCT band movement analysis. The intensity-based processing takes advantage of the fact that morphophysiological changes in the retinal neurons and vasculature can result in local variation of IOS properties, such as refractive index, scattering, reflection, or birefringence.

#### 3.1.1 OCT brightness analysis

The OCT brightness change analysis was devised to detect local variations in pixel intensity value due to light stimulus within the retina. Previous studies using brightness change analysis detected localized IOS change both in the inner and outer retina [42; 51; 58; 59; 60]. **Figure 2** illustrates representative time-lapse OCT recording and a result from the brightness change analysis [61]. The data processing is described as follows [42; 60; 62]: Raw OCT B-scans are first registered to compensate eye movements using a sub-pixel registration algorithm. Each pixel intensity is then normalized relative to the inner retina before brightness change analysis to minimize the effect of pupillary response [25]. Next, the 'active' IOS pixels are identified from a sequence of the registered OCT B-scans. Any pixel that markedly changes its value in response to the light stimulus is identified as the active-IOS pixel. To avoid random speckle noise, two conditions are sequentially applied: 1) the three-sigma (3-$\sigma$) rule and 2) the consecutive rule, i.e., a pixel that changes its intensity value greater or less than pre-stimulus mean $\pm$ 3 standard deviations (3-$\sigma$) over the consecutive number of the post-stimulus frames is assigned as an active IOS pixel, which can be either positive or negative sign. The number of active IOS pixels can be quantified for comparative study [39; 62]. In addition, the intensity value of active IOS pixels can be traced over time after subtracting the background pixel intensity value from pre-stimulus B-scans [59; 61]. **Figure 2** demonstrates that light-evoked positive (red) and negative (green) IOSs were observed in the human retina. As shown in **Figure 2(B)-(D)**, the fast IOS changes was promptly observed after the onset of stimulus and primarily confined within the photoreceptor region. Because ~220 ms time window was available for IOS imaging without the effect of pupillary response (**Fig. 1B2**), it would be feasible to monitor the fast photoreceptor-IOS in non-mydriatic condition [25].

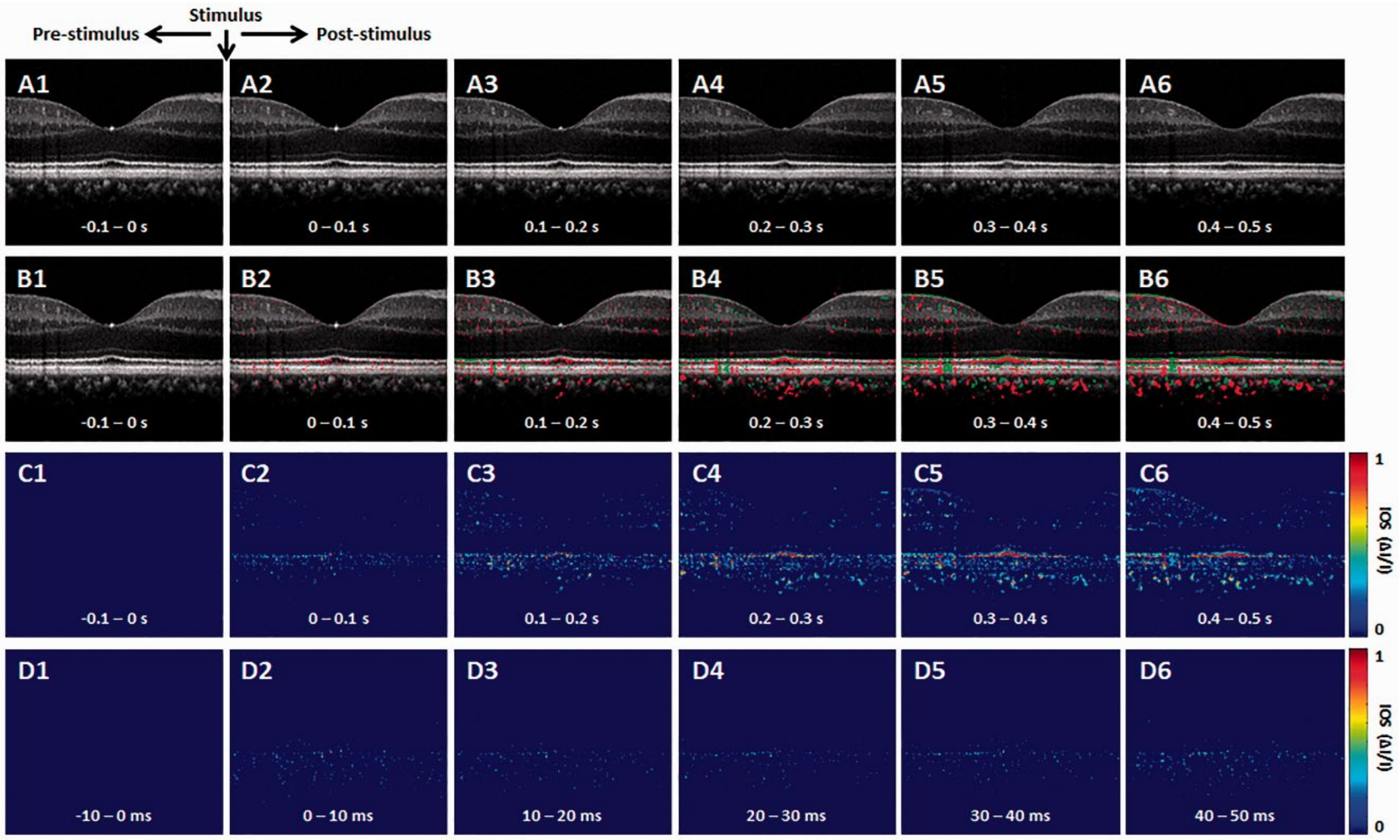

**Figure 2.** Representative functional OCT intensity-based processing IOS/ORG imaging. (A) Representative OCT image sequence. (B) Corresponding IOS distributions of positive (red) and negative (green) changes. (C) IOS magnitude sequence with 0.1 s time intervals. (D) IOS magnitude sequence with 10 ms time intervals. Reprinted with permission from Son et al [25].

### 3.1.2 OCT band analysis

The retina is stratified into multiple-layered structures. OCT can probe the axial position of each neural and synaptic layer and visualizes these layers as hyper- and hypo-reflective bands. Measuring the band alternations under different light conditions is one of key parameters in ORG measurement [35; 54; 63; 64; 65]. There is a growing interest especially in assessing outer retinal bands, such as photoreceptor inner segment (IS), OS, and subretinal space (SRS), and retinal pigment epithelium (RPE).

**Figure 3** illustrates outer retinal bands and analysis methods. Zhang et al. demonstrated a deconvolution method for band analysis (**Fig. 3A**). OCT spectra were first Fourier transformed after 4-times zero-padding, and the resultant A-scan profiles remapped onto a linear scale. Next, they used deconvolution to extract additional information from the hyperreflective bands. Specifically, the averaged A-line profiles were deconvolved with the MATLAB “deconvlucy” function, and the hyperreflective bands from each time point fitted with Gaussian functions, providing three parameters (position, amplitude, and full width at half maximum) that can be used for OCT band analysis. This method was used to measure the length of photoreceptor OS over the diurnal cycle in albino mice [35]. Messner et al. demonstrated modeling the OCT A-line profiles by fitting normal distribution curves and observing their position changes over time (**Fig. 3B**). In order to better determine the position of outer retinal bands, a signal model for the A-scan averages was developed in the software OriginPro 2019b using the “multiple peak fitting” function. The signal model for the A-scan average allowed for tracking the position of the peaks attributed to the boundaries of the outer retinal layers during baseline conditions and stimulation of the retina, providing quantitative parameters [54]. In addition, Kim et al. recently demonstrated transient band shifting during the initial dark adaptation period in the mouse retina. The high-speed recording captured 4800 B-scans at the same retinal plane with 16 frames per second rate for a 5-minute recording. Volumetric average was conducted for OCT A-line band analysis, and linear interpolation was employed to enhance the sampling density of average A-line band analysis [63]. Yao et al. suggested more detailed band analysis by accounting for not only hyper-reflective band location but also relative distances between hyper- and hypo-reflective bands to better establish the correlation of each band to the outer retina structure. This approach may provide additional insight into the outer retinal structure and its dynamics under different light conditions [66].

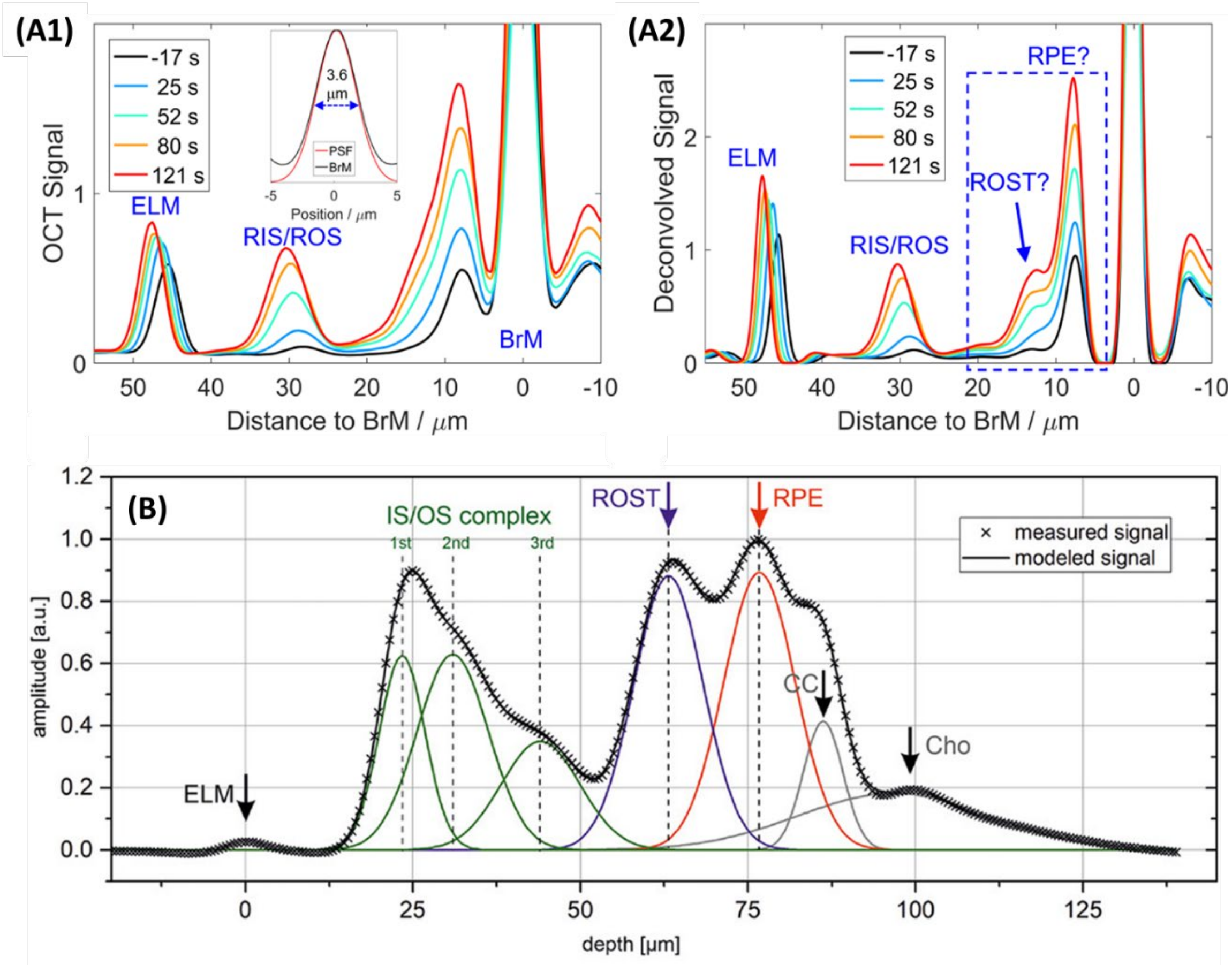

**Figure 3.** Representative OCT hyper-reflective band analysis. (A) Deconvolution method. (A1) Depth scattering profiles of the retina of an albino mouse. (A2) Deconvolution analysis reveals that the backscatter band nearest to Bruch's membrane (BrM) on the anterior side comprises two distinct components (question marks are used to indicate that the assignment to structures required confirmation). Reprinted with permission from Zhang et al [35]. (B) A-line signal modeling by summation of seven Gaussian curves. Gray crosses represent the measured data points, and the black line is the summation of the individual model curves (green, blue, red, and gray lines). ELM: external limiting membrane; IS/OS: inner segment/outer segment complex; ROST: rod outer segment tips; RPE: retinal pigment epithelium; CC: choriocapillaris; Cho: choroid. Reprinted with permission from Messner et al [54].

## 3.2 OCT phase-based processing

Evaluation of the phase of interference fringes allows access to information about OPL changes. Recent development of phase-resolved OCT offers sensitivities to photoreceptor OS deformation on a nanometer scale, much smaller than the axial resolution of the OCT system.

### 3.2.1 OCT band boundary measurements for optical path length estimation

Given the premise that the photoreceptor OSs change in length by light stimulus, OCT phase information has been used to estimate the OPL change of photoreceptor OSs. The photoreceptor OS is long and narrow; thus it can behave like an optical waveguide [67]. In addition, there are relatively strong reflections from each end of the photoreceptor (IS/OS junction and OS tip), well suited for OPL estimation using hyper-reflective band positions. Recent advance in parallel OCT and incorporated AO subsystem enables to measure stable phase information. Since the phase in single layer does not carry information to evaluate length change of the photoreceptors, it is necessary to compare two phases between two different retinal layers and between two different time points. **Figure 4** demonstrates phase-resolved OCT imaging for OPL estimation. The data processing is described as follows [55]. First, the recorded OCT volumes were reconstructed, and each pixel of each reconstructed volume was first referenced to the respective co-registered pixel in one specific volume. Next, the layers and pixels that carry the information about the OS length need to be segmented. Two layers used for segmentation are generally photoreceptor OS tips (POST) and inner-outer segment junction (IS/OS). In general, several axial pixels are averaged centered around peak point of each layer. The temporal evolution of optical phase difference is then computed between the POST and IS/OS ($\Phi_{POST}$ - $\Phi_{ISOS}$) to yield a measure of light-induced relative phase changes between POST and IS/OS. The phase difference at the two layers is then converted to OPL using the relation $\Delta OPL = (\lambda_c/4\pi) \times (\Phi_{COST} - \Phi_{ISOS})$, where $\lambda_c$ = central wavelength of OCT light source. A study showed that the magnitude of these OPL changes was strongly correlated with light-induced activity, and they utilized this correlation to classify three cone classes [68]. In principle, relative OPL changes between any two layers including inner retinal layers can be estimated [69]. The stability of phase data is crucial in the OPL estimation and often corrupted by ocular movements. Thus, the phase-resolved ORG measurement generally requires ultrahigh-speed recording and voxel-wise registration. It should be also noted that different shapes of cone OSs presented from the central fovea to the parafovea may affect OPL estimation [70].

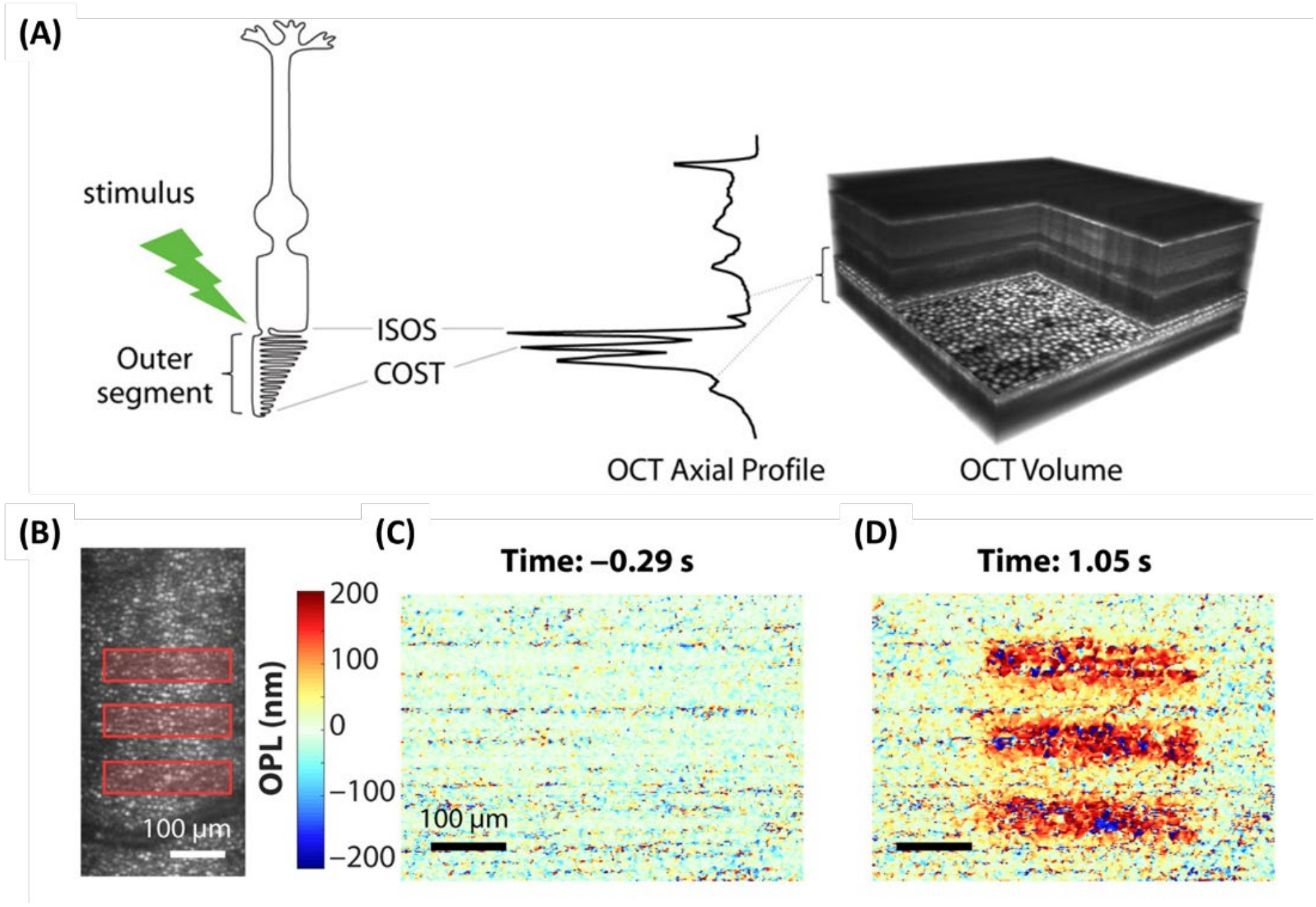

**Figure 4.** Phase-resolved OCT imaging for optical path length (OPL) estimation. (A) Optoretinography experimental paradigm. A three-dimensional (3D) OCT volume with AO allows resolving the cone mosaic in an en face projection and the outer retinal layers in an axial profile corresponding to the ISOS and COST. Stimulus (528 ± 20 nm, green)–driven changes in a cone photoreceptor are accessible by computing the time-varying phase difference between the proximal and distal OCT reflections encasing the outer segment. (B) Optoretinography reveals functional activity in cone outer segments. Illumination pattern (three bars) drawn to scale over the line-scan ophthalmoscopic image. (C and D) The spatial map of OPL change between the ISOS and COST before (C) and after stimulus (D), measured at 20-Hz volume rate. Reprinted with permission from Pandiyan et al [32].

### 3.2.2 Differential-phase analysis for spatiotemporal mapping at pixel resolution

Ma et al. recently demonstrated a new approach that can simultaneously monitor the phase changes along the whole retinal depths, called differential phase mapping (DPM) [37]. DPM was devised to analyze the spatiotemporal phase change at pixel resolution. For DPM processing, the OCT phase is unwrapped followed by differentiation along the A-scan direction (**Fig. 5A**). If the scatters of the adjacent pixels are both in the center, the value of the pixel in DPM is $4\pi nL/\lambda_c$, where $n$ denotes the refractive index of the tissue, $L$ denotes the pixel length, $\lambda_c$ denotes the center wavelength of the light source, and the coefficient $4\pi$ is due to the OCT measures the back scatter light. If the distance of two scatterers is less than $L$, the pixel value of DPM will be smaller than $4\pi nL/\lambda_c$, vice versa. Therefore, the DPM represents the relative scatter distance of the sample. Compared to the OCT amplitude image (**Fig. 5B**), DPM also reveals the structural information in the representation of scatter locations (**Fig. 5C**). After stimulation, both amplitude and phase IOS appeared at the outer retina layers (**Fig 5D and 5E**). The phase IOSs of different layers at different time courses indicated the depth association of phototransduction in the outer retina. Compared to conventional phase resolved OPL measurement, computing the phase change between two selected locations, DPM shows the phase change over all the retina depths simultaneously, which could help to understand the phase dynamic between retinal layers.

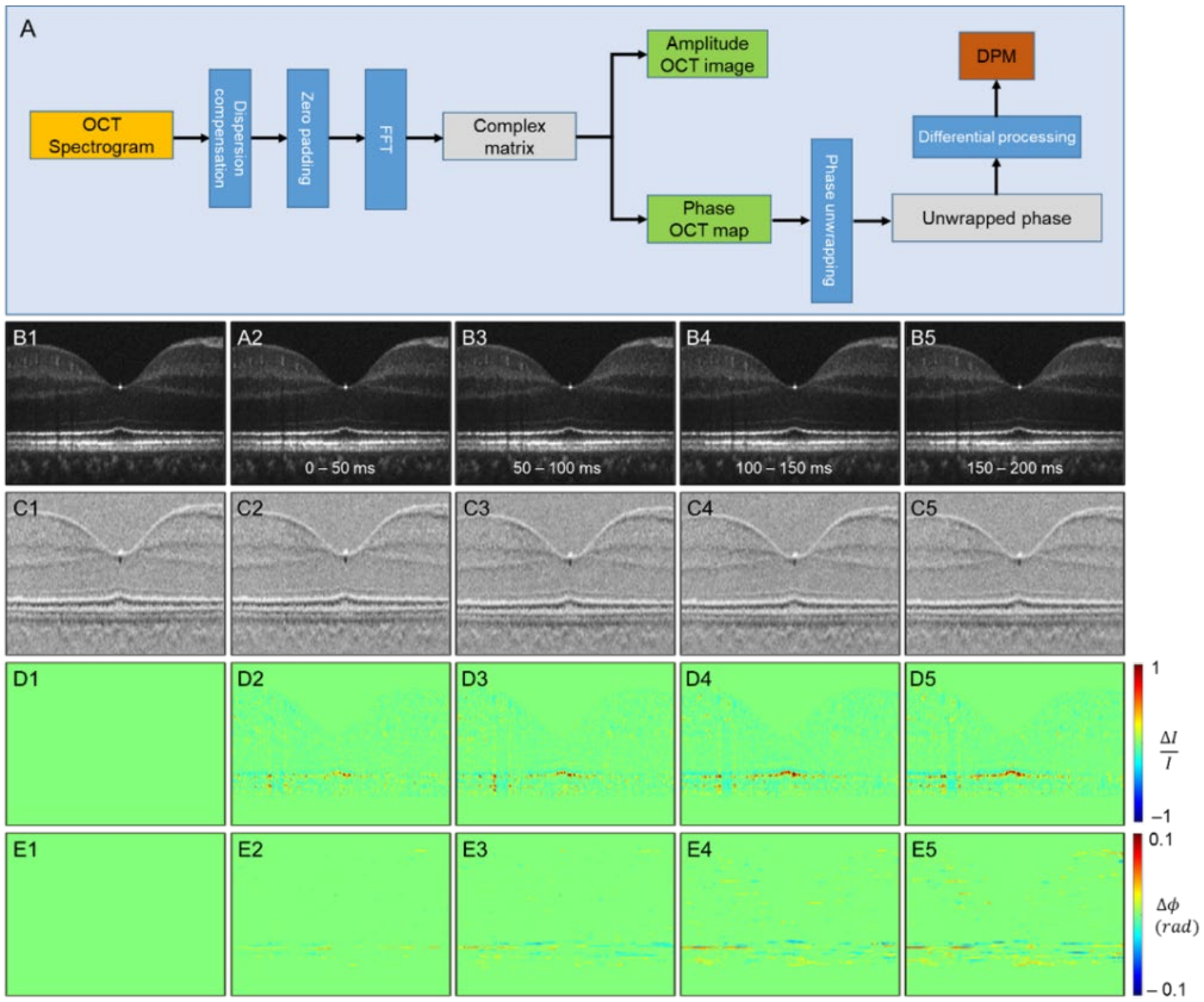

**Figure 5.** Phase-resolved OCT imaging for differential-phase mapping (DPM) analysis. (A) A flow chart of amplitude OCT and DPM processing. (B-E) The amplitude-IOS and phase-IOS distribution. (B) Amplitude image sequence. (C) DPM sequence. (D) Amplitude-IOS sequence. (E) Phase-IOS sequence. Reprinted with permission from Ma et al [37].

## 4 ORG interpretations and its challenges

ORG measurement and interpretation is challenging due to the multiple signal sources and variable OCT instruments and experimental protocols.

### 4.1 Retinal neurovascular coupling and inner retinal response

Retinal blood flow is actively regulated in response to neuronal activity [71], called neurovascular coupling. Impaired coupling mechanism is commonly associated with microvascular pathologies in the retina [72; 73]. Thus, spatiotemporal mapping of transient neural activity and subsequent hemodynamic responses promises early detection of retinal diseases. Based on the intensity-based processing, stimulus-evoked IOS changes have been observed in both retinal and vascular layers. Son et al. demonstrated neural- and hemodynamic-IOS are typically mixed in the retina and reported concurrent mapping of neural-IOS and hemodynamic-IOS changes to enable functional monitoring of retinal neurovascular coupling [50; 61]. They leveraged OCT angiography (OCTA) map to isolate the retinal vasculature at a single capillary level resolution. The OCTA-guided IOS data processing enables two functional images: a neural-IOS map and a hemodynamic-IOS map. Flicker stimuli was used to induce robust hemodynamic response. As shown in **Figure 6**, fast photoreceptor-IOS change was first observed right after the stimulus onset; while hemodynamic-IOS revealed with a significant time delay, compared to the rapid photoreceptor-IOS. Different time courses and signal magnitudes of hemodynamic-IOS responses were also observed in blood vessels with various diameters (**Fig. 6D**). However, the mechanism for the different hemodynamic-IOS responses of large blood vessels and small capillaries is not understood yet. Only a few hypotheses have been proposed such as different neural metabolic demands in individual retinal layers, passive dilation of downstream capillaries, and mural cells' intervention on blood flow regulation [61]. In addition, 2D cross-sectional imaging would be challenging for monitoring delayed hemodynamic IOS in the human retina due to motion artifacts. Moving correction is impossible if the retina moves perpendicularly to the imaging plane as there is no data to use in correction. High-speed parallel OCT would be desirable for neurovascular coupling study. Another task is to appreciate how local signal variations occur in blood vessel regions associated with flux change, hematocrit, and diameter increase.

Aside from hemodynamic information, the inner retina itself also revealed IOS change (**Fig. 6C**). Previous mouse studies reported that a short 10-ms stimulation mainly induced the fast-photoreceptor IOS changes, while increased stimulus duration or flicker stimulation induced IOS changes in the inner retina [41; 42; 50]. Although the signal source is underappreciated, it was postulated that the slow inner retinal IOS might be associated with an integral effect of electrophysiological signal transduction between multiple inner retinal neurons, such as horizontal cells, bipolar cells, amacrine cells, and ganglion cells. The plexiform layer in the retina consists of complex synaptic network containing numerous dendrites from different types of neurons [74; 75]. Their synaptic signaling might affect optical signal properties. Pfäffle et al. recently showed simultaneous imaging of the activation in the photoreceptor and ganglion cell layer/inner plexiform layer (GCL/IPL) in the human retina by using phase-sensitive full-field swept-source OCT (FF-SS-OCT) [69]. Although the signals from the GCL/IPL were 10-fold smaller than those from the photoreceptor, GCL/IPL signal were still detectable with suppression of motion artifacts and pulsatile blood flow in the retinal vessels. The phase difference of the GCL and the IPL was calculated to evaluate the light evoked OPL changes, and they found that the OPL between GCL and IPL increased about 40 nm in the stimulated area, and the increase in OPL reached its maximum of about 40 nm after approximately 5 s. However, the mechanism of inner retinal IOS changes from both intensity-based and phase-based results is poorly understood. In addition, retinal vasculature is embedded in the inner retina; thus, blood flow pulsation and inhomogeneous intensity distribution may complicate signal interpretations.

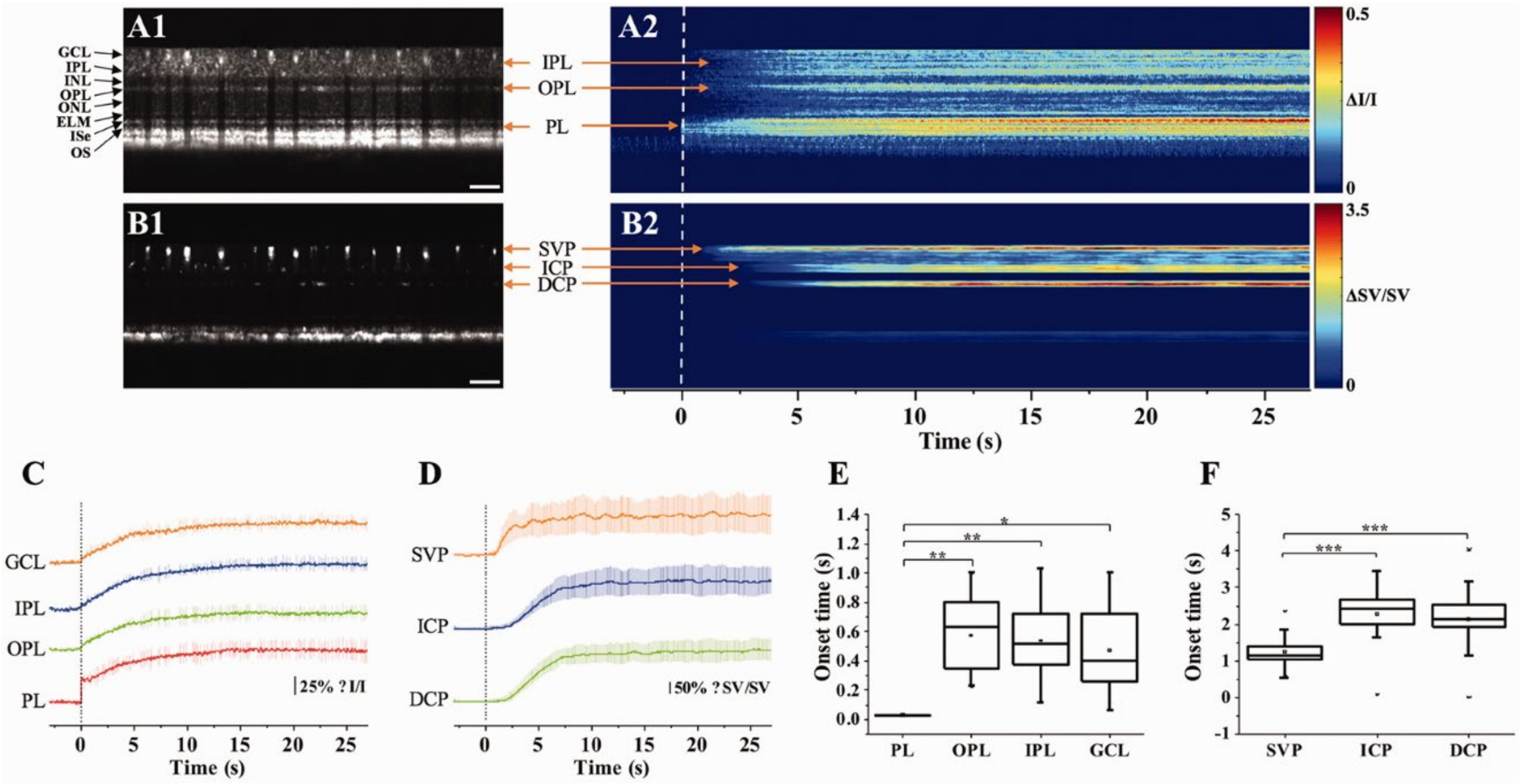

**Figure 6.** Retinal neurovascular coupling and inner retinal IOS response. (A) Representative flattened (A1) OCT B-scan and (A2) spatiotemporal neural-IOS map. (B) Representative flattened (B1) OCTA B-scan and (B2) spatiotemporal hemodynamic-IOS map. Scale bars in (A1) and (B1) indicate 500 µm. (C) Neural-IOS changes of photoreceptor layer (PL), outer plexiform layer (OPL), inner plexiform layer (IPL), and ganglion cell layer (GCL). (D) Hemodynamic-IOS changes of superficial vascular plexiform (SVP), intermediate capillary plexiform (ICP), and deep capillary plexiform (DCP). (E) Averaged onset times of neural-IOS changes at PL, OPL, IPL, and GCL. (F) Averaged onset times of hemodynamic-IOS changes of SVP, ICP, and DCP. Modified with permission from Son et al [61].

## 4.2 Transient deformations of photoreceptor outer segment

As the center of phototransduction, retinal photoreceptors are responsible for capturing and converting photon energy to bioelectric signals for following visual information processing in the retina. Retinal photoreceptors are the primary target cell of retinal degenerative diseases such as AMD and retinitis pigmentosa (RP); thus, noninvasive monitoring of functional integrity of photoreceptors is of great interest. The photoreceptors are the most well studied among retinal cell types by intrinsic signal ORG measurement as they exhibit an exceptionally reproducible light-driven response.

Time-lapse NIR light microscopy was initially used to image transient IOS changes in freshly isolated retinas, and it was found that the IOS change rapidly occurred in the photoreceptor cells after the visible light stimulus [76; 77]. The IOS magnitude and time course were found to be dependent on the stimulus strength [76]. In addition, transient photoreceptor deformation was directly observed in both amphibian [78] and mammalian [79] retinas. The transient deformation was shrinkage-induced and predominantly observed in the photoreceptor OSs that rapidly shifted toward the direction of the visible light stimulus [47]. It turned out that the onset time of transient photoreceptor shrinkage was almost identical to that of photoreceptor-IOS change. This study further suggests that the OS conformational change may correlate with phototransduction process. Dynamic confocal microscopy and OCT study further suggested that the photoreceptor OS is the anatomic source of the transient photoreceptor movement [47]. Lu et al. showed vertical shrinkage of isolated frog rod OSs by a visible light stimulation, and transmission electron microscopy (TEM) observation further confirmed shortened inter-disc spacing in light-adapted rod OSs compared to that in dark-adapted rod OSs [80]. To better understand the physiological mechanism of the fast-photoreceptor IOS, comparative measurements of OS movement and ERG were conducted. It was consistently observed that the conformational change of OSs occurs earlier than the onset of the ERG a-wave that reflects the hyperpolarization of retinal photoreceptors [81]. Moreover, substitution of normal superfusing medium with low-sodium medium reversibly blocked the photoreceptor ERG a-wave, but largely preserved the stimulus-evoked rod OS movements [81]. In comparative photoreceptor-IOS recording and ERG measurement, previous studies confirmed the response time of fast photoreceptor-IOS was clearly ahead to the a-wave in the mouse retina [39; 60]. This observation provides solid evidence that the fast-photoreceptor IOS does not depend on OS plasma membrane hyperpolarization, i.e., cyclic guanosine monophosphate (cGMP) gated ion channel closure, which is the source of ERG a-wave. Rather, the fast IOS and rod OS conformational changes are tied directly or indirectly to the early phase of phototransduction process that involves the sequential activation of rhodopsin, transducin, and cGMP phosphodiesterase (PDE). A recent comparative study of wild-type (WT) and retinal degeneration 10 (rd10) mice demonstrated that fast photoreceptor-IOS occurs even earlier than PDE activation [82]. Similarly, recent phase-resolved OCT imaging revealed stimulus-evoked rapid reduction of OPL in photoreceptor OSs in the human retinas [32; 56; 68]. The rapid OPL decrease showed a time course of millisecond level, which is consistent with that of stimulus-evoked fast photoreceptor-IOS in animal models [45]. Boyle et al. suggested that contraction of the photoreceptor OS may be driven by the charge transfer across the OS disc membrane relevant to early receptor potential (ERP) [32; 83], a fast electrical signal observed in cone photoreceptors under intense flash stimuli [84; 85]. The ERP is associated with the conformational change of opsins embedded in the OS disc membrane and distinct from the later changes in the photoreceptor membrane potential.

Intriguingly, phase-resolved OCT imaging revealed not only a rapid (<5 ms) reduction in OPL after the stimulus onset, but also a slower (>1 s) increase in OPL of the photoreceptor OSs [31; 32; 56; 68]. The elongation response has been consistently observed in phase-resolved ORG studies. Zhang et al. showed that the magnitude of these path length increment was positively correlated with stimulating light dose, and they used the photoreceptor elongation signals to generate maps of the three cone classes [68]. It has been hypothesized that the increase (up to hundreds of nanometers) in the OPL of photoreceptor OSs after light stimulus would be attributable to osmotic swelling, an increase in the cytoplasmic volume due to excess osmolytes produced by phototransduction [64]. Based on the intensity-based processing, Lu et al. also observed photoreceptor OS elongation following strong light exposure and subsequent recovery, i.e., photoreceptor OS shortening, in human subjects [86]. Although the increased OPL between IS/OS junction and OS tip is consistent among *in vivo* phase resolved ORG measurement, there is a lack of direct evidence of the OS elongation. In fact, *ex vivo* studies have often showed conflicting results (**Fig. 7**). Comparative TEM has revealed shortened inter-disc spacing of OSs in light-adapted retinas compared to that in dark-adapted retinas [80]. Fast X-ray diffraction studies also found a light-induced shrinkage of the disc lattice distance from both the frog and mouse rod OSs [87; 88]. Moreover, Bocchero et al. recently measured light-evoked 3-axis (X, Y, Z plane) volume changes in the single rod OS from Xenopus retina [89]. They consistently observed a shortening of the OS on the order of 100–200 nm after a brief flash stimulus. The shortening was transient, and the OS returned to its original size within about 10 s, without further expansion.

Taken together, there is a growing consensus on the photoreceptor OS shrinkage at the early stage of phototransduction. However, more research is necessary to verify the OS swelling or relaxation mechanism. Note that the different dynamics between retinal explant and isolated single photoreceptor responses were demonstrated [87], and fundamentally, photoreceptor morphology and cellular compartment are different among species [90; 91].

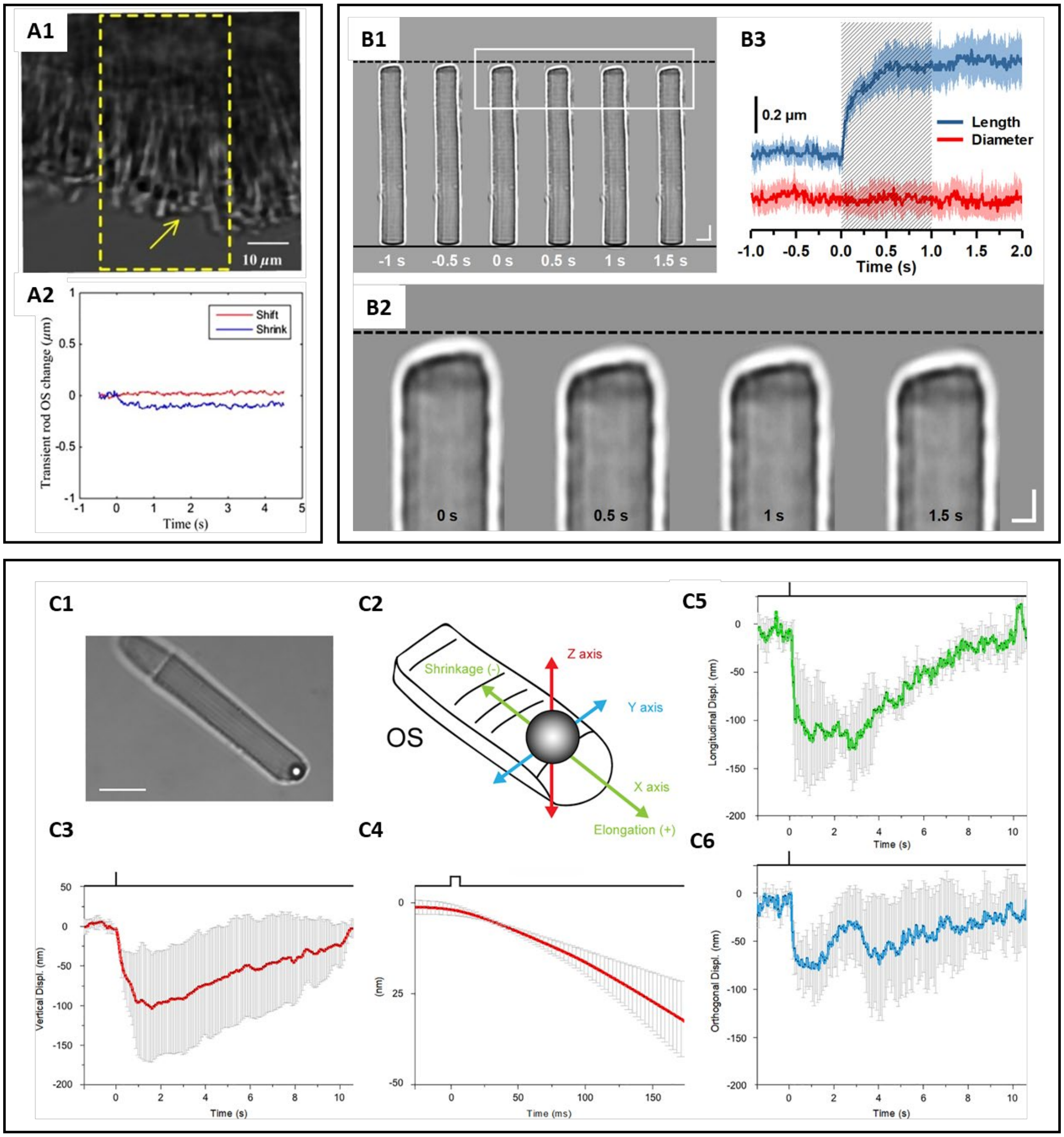

**Figure 7.** Photoreceptor outer segment (OS) shrinkage due to light stimulus. (A1) Stimulus-evoked mouse rod OS movement. The yellow window indicates the stimulation area. (A2) OS changes of a mouse rod photoreceptor. In the center of the stimulation region, the length of the OS shrunk, while in the peripheral region, the OS swung toward the center of the stimulation area in the plane perpendicular to the incident stimulus light. Reprinted with permission from Zhao et al [92]. (B1) Representative light microscopic images of an isolated frog rod OS acquired with an interval of 0.5 seconds. To better show the light-evoked OS shrinkage, the base of the rod OS in each image is aligned horizontally as shown by the black solid line at the bottom. The black-dashed line at the top represents the position of the rod OS tip at time −1 second. Scale bars (in white) represent 5 µm. (B2) Enlarged picture of the white rectangle in B1. Scale bars (in white) represent 2 µm. (B3) Time course of the averaged rod OS shrinkage in both length and diameter acquired from eight different rod OSs. Colored areas accompanying the curves represent the standard deviations. Shaded area indicates the 1-second stimulation period. Reprinted with permission from Lu et al [80]. (C) Mechanical response of an X. laevis rod to light flashes. The position of a bead sealed against the tip of the rod OS is monitored with optical tweezers. Following a bright flash of 491 nm, equivalent to about $10^4$ $R^*$, a transient shrinkage is observed. (C1) Bright-field infra-red image, showing a trapped bead in contact with the tip of the rod OS (scale bar, 10 µm). (C2) Detail of the 3D tracking system. (C3) Light-induced shifts in the Z axis of the trapped bead (downward is negative). (C4) Expansion of the time base in C5 to examine the delay between light stimulus and bead movement. (C5) Bead displacement along the direction of the rod OS (shrinkage is negative, and elongation is positive). (C6) Bead displacement in the direction perpendicular to the rod OS axis. Data are representative of mean ± SD of 5 different experiments. Reprinted with permission from Bocchero et al [89].

### 4.3 Transient reflectance changes in inner segment ellipsoid zone

The photoreceptor ISe is the center of metabolism, consisting of abundant mitochondria [93]. Despite of contradictory nomenclatures, the ISe is a widely used biomarker of photoreceptor structure. The integrity of the ISe band, lesion size, and width of retained ISe are established metrics that have been correlated with visual acuity and other aspects of retinal function [94]. More recently, ISe reflectivity has emerged as a sensitive biomarker of photoreceptor structure, because ISe reflectivity has been shown to undergo changes in retinal degenerative conditions before marked changes in ISe integrity [94]. In addition, recent IOS recording showed that the ISe reflectivity dynamically changed in response to different light conditions [59; 63; 64]. **Figure 8** shows the ISe IOS change in the mouse retina. A study demonstrated that stimulus-evoked IOS change at the ISe appeared with a time delay of ~12 ms after the stimulation, which is rather slower than fast-OS response, suggesting that the slow ISe IOS might reflect the metabolic reaction of mitochondria, following the phototransduction in the OS [59]. Under metabolic stresses, the morphological structure, motion dynamics, and fission or fusion configuration of mitochondria are all varied [95; 96], which could alter the optical signal properties of the ISe zone, resulting in OCT reflectance changes. In addition, Kim et al. recently found a significant reduction in ISe reflectance during dark adaptation in the mouse retina. This observation further emphasizes light-modulated ISe reflectivity could serve as a sensitive biomarker for photoreceptor dysfunction [63]. However, P. Zhang et al. hypothesized that phototransduction reactions associated with complete activation of G-protein alpha-subunit transducin may induce osmotic swelling, and the swelling, combined with mass redistribution of transducing proteins from the disc membranes into cytosol, might in turn underlie the scattering increases at the IS/OS and OS tips [64]. There is no doubt that the ISe reflectivity can be actively regulated under different light conditions, and it would be a potent biomarker for photoreceptor dysfunction. However, the ISe IOS source is still ill-defined, and both mitochondrial metabolic activity and redistribution of G proteins could simultaneously affect the signal. In addition, there is unmet need to resolve ambiguity as to how well ISe reflectivity correlates with underlying photoreceptor structure [66].

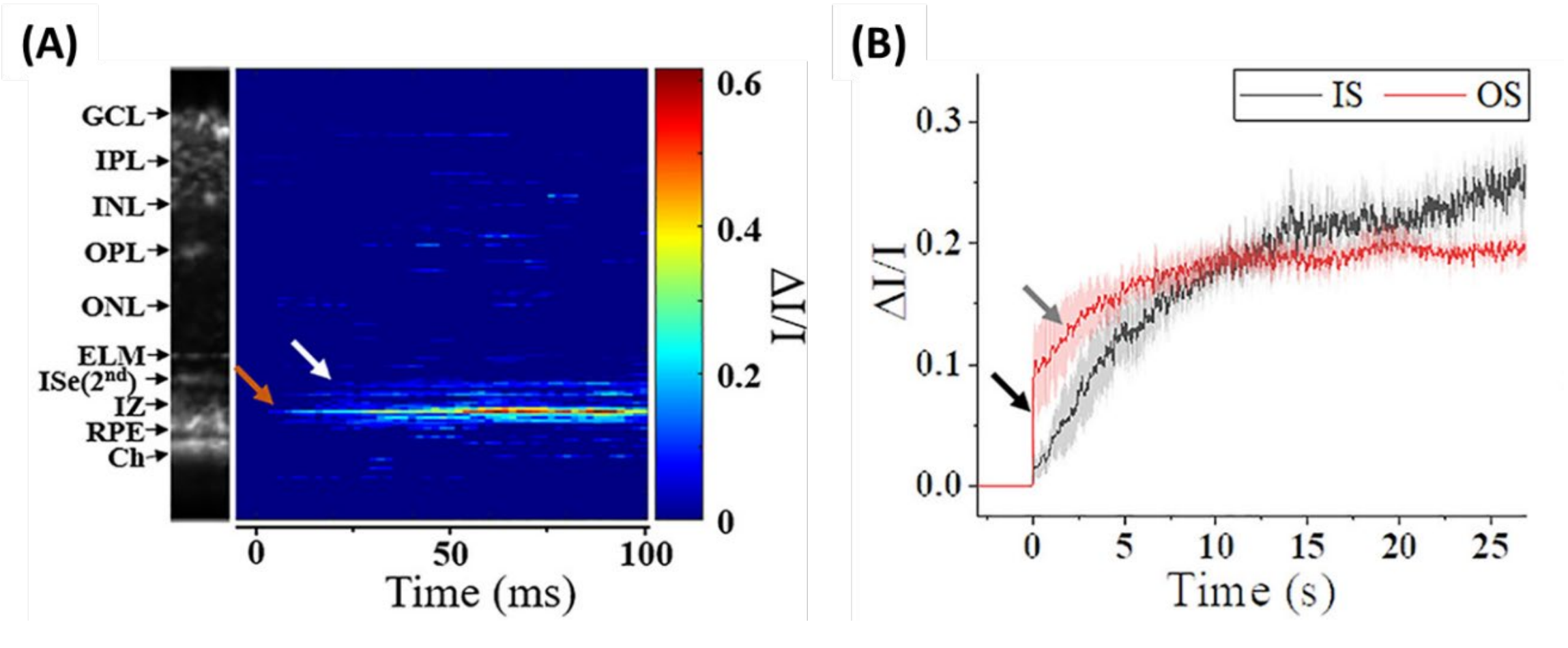

**Figure 8.** Metabolic response of photoreceptor inner segment. (A) IOS M-scan within 100 ms, the white and red arrowheads show the onsets of IS-IOS and OS-IOS, respectively. (B) Average IS-IOS and OS-IOS from six mice. The average OS-IOS showed a biphasic curve. The dark and gray arrowheads show the first rapidly increasing phase and the second gradually increasing phase, respectively. GCL, ganglion cell layer; IPL, inner plexiform layer; INL, inner nuclear layer; OPL, outer plexiform layer; ONL, outer nuclear layer; ELM, external limiting membrane; ISe (2nd), inner segment ellipsoid (the 2nd

hyper-reflective band); IZ, interdigitation zone; RPE, retinal pigment epithelium; Ch, Choroid; IS, photoreceptor inner segment; OS, photoreceptor outer segment. Reprinted with permission from Ma et al [59].

## 4.4 Subretinal space changes under different light conditions

The SRS is the extracellular fluid space between photoreceptors and RPE cells, spanning from the ELM to the apical RPE. This space is isolated by tight junctions at these two borders, and the ELM and RPE appear as hyper-reflective bands in OCT, which facilitates band change analysis. In fact, the SRS has been well recognized region that actively deforms under different light conditions [97]. Thus, SRS dynamics could become a potential biomarker for outer retinal dysfunction.

Using Fourier-domain OCT with intensity-based processing, Li et al. found a significant reduction of outer retinal thickness in the dark-adapted mouse retina [98]. They also reported that light-dependent volume changes in the outer retina varied with the stage of retinal degeneration in retinal degeneration 10 (rd10) mouse model [65]. Berkowitz et al. found strain-specific changes of the outer retina. A light-driven expansion of the outer retina was more prominent in C57BL/6 mice than 129S6/SvEvTac mice [99]. Gao et al. further demonstrated that dark adaptation significantly reduced the magnitude and width of a hypo-reflective band between the photoreceptor OS and RPE in the mouse and human retina [100]. Lu et al. also showed a rapid decline in the IS/OS-RPE distance after a light stimulus [86]. **Figure 9** illustrates the dark adaptation effects on the mouse retina. Kim et al. recently demonstrated the dynamic SRS thinning of the mouse retina during light-dark transitional moment, and found that retinal response during dark adaptation was reflected by transient structural (i.e., the SRS thinning) and physiological (i.e., ISe intensity reduction) changes [63]. High-speed OCT recording further identified a strong correlation between the SRS thinning and ISe intensity reduction in the outer retina [63]. The SRS thinning mechanism is rather well understood by RPE-mediated water removal from the ELM-RPE region. The transition from light to dark is accompanied by an increase in photoreceptor metabolism, resulting in increased oxygen consumption in the retina [101], which can acidify the outer retina due to increased CO2 and wastewater production, and ultimately upregulate water removal co-transporters in the RPE. This rapid removal of the acidified water has been linked to significant thinning of the ELM-RPE region [97; 102].

However, human study has sometimes showed intriguing but puzzling results. Messner et al. demonstrated distance decrease between IS/OS junction and RPE after light stimulus [103]. They observed that the OS tips and RPE were drawn closer together after light exposure, speculating that this might be associated with a decrease in volume of the SRS [103], which is conflicting with previous observations [97]. In addition, Azimipour et al. observed that the RPE band appears to split, with its apical portion moving toward cone OS tip after light stimulus. They speculated that light-driven translocation of melanin observed in amphibians [49] or a consequence of inward water movement across the RPE/Bruch’s complex could be a potential signal source [104]. In the human retina, photoreceptor cell population, rod/cone ratio, photoreceptor length, and morphology are largely different depending on eccentricity. Appearance of outer retinal bands also varies among different OCT system [66]. Thus, various factors that can potentially affect ORG measurement should be carefully accounted and documented.

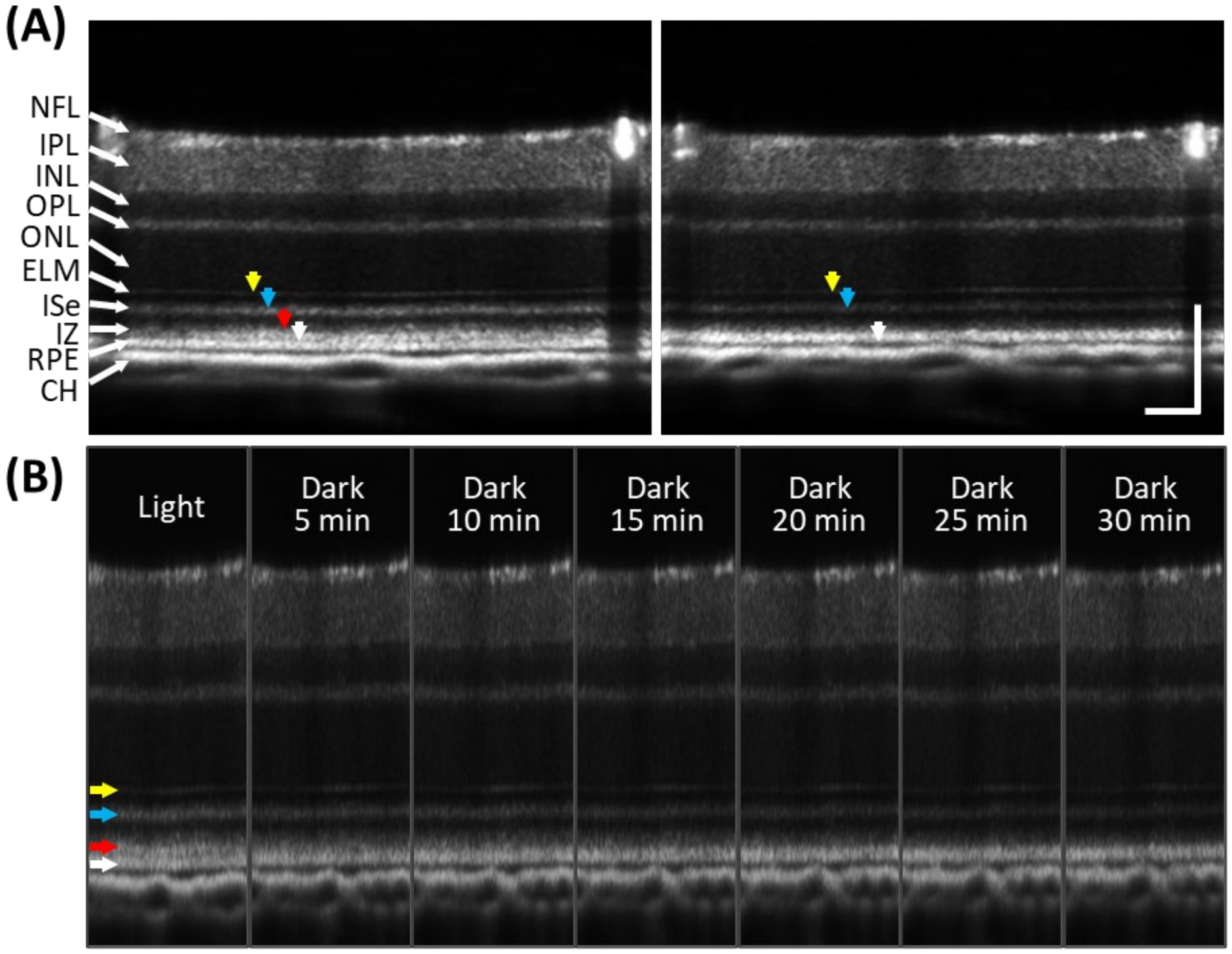

**Figure 9.** Subretinal space changes during dark adaption in mouse retina. (A) OCT images of light- and dark-adapted retina of a two-month-old C57BL/6J mouse. Color arrows to indicate outer retinal bands: 1st ELM band (yellow), 2nd ISe band (blue), 3rd IZ and OS tip band (red), 4th RPE band (white). (B) A sequence of OCT images obtained every 5 min up to 30 min during dark adaptation. During dark adaptation, ISe intensity reduction rapidly occurred, and the SRS became thinner. In addition, the 3rd outer retinal band (red arrow) faded over time. NFL: nerve fiber layer; IPL: inner plexiform layer; INL: inner nuclear layer; OPL: outer plexiform layer; ONL: outer nuclear layer; ELM: external limiting membrane; ISe: inner segment ellipsoid; IZ: interdigitation zone; RPE: retinal pigment epithelium; CH: choroid. Scale bars: 100 µm.

# 5 Discussion

## 5.1 Emerging issues in ORG measurements

As demonstrated above, outer retinal structures including the photoreceptor IS and OS, SRS, and RPE have been highlighted in ORG recording because of their structural clarity in OCT and importance as a primary target of retinal degenerative diseases. However, ORG recording has been conducted by different OCT systems and experimental protocols. Thus, there are often challenges for quantitative analysis using OCT reflectance information. Lee et al. recently provided a valuable discussion about challenges associated with ISe intensity measurement using clinical OCT [94]. They emphasized the importance of pupil entry position of the OCT beam due to altered reflectivity profile of the retinal image [105; 106]. Without standardization of pupil entry point acquisition, the reproducibility of ISe reflectivity measurements will be limited. There is also the importance of using appropriate intensity scale. Clinical OCT images are generally presented in a logarithmic scale, but this can result in misrepresentation of real differences in reflectivity and a loss of information [107; 108]. By adjusting gray values, hyper-reflective outer retinal bands can be broadened and their vertical position can be altered within the scan [107]. Another issue is inter-device variation in ISe intensity [109]. It has been shown that normalization of ISe reflectivity as a ratio of the intensity of the ISe band to a retinal layer allows for comparison across subjects, devices, and time points. However, this can be a complicating factor since each OCT device has different methods of image acquisition and optimization. Thus, it is necessary to establish a standardized normalization method [94]. Meleppat et al. further demonstrated that the reflectivity from the inner retinal layers and ELM are also highly directional, as their strength declines steeply with the angle of incidence of the OCT-scanning beam on the mouse retina [110]. Notably, in albino mice, the reflection from Bruch's membrane was highly directional as well [110].

Aside from the issues of image acquisition and analysis, the unmet need for advancing ORG measurement is to establish anatomic correlation with underlying photoreceptor structure. Yao et al. recently provided a valuable discussion about interpretation of anatomic correlates of outer retinal bands in OCT [66]. Either clinical OCT or AO-OCT enables to resolve 4 distinct hyperreflective bands in the outer retina. Recent resolution improvement allows further separation of the fourth band into the RPE and Bruch's membrane [74]. However, our understanding of anatomical correlates to each outer retinal band is not keeping pace with the recent development of OCT imaging technology. Understanding anatomical correlates to OCT bands is crucial to translate ORG information for diagnostic purpose. The interpretations of the first and fourth bands of the outer retina are consistent, but the second and third bands remain a great controversy [107; 111; 112]. Particularly the inconsistency is generally observed in between clinical OCT and AO-OCT. Comparative alignment of the outer retinal OCT bands with an anatomically correct model suggested the ISe and interdigitation zone (IZ) as the correlates to the second and third bands [113], and the 2014 International OCT Nomenclature Meeting affirmed the ISe and IZ as the anatomic correlates of the second and third bands [114]. However, with improved spatial resolution, AO-OCT study revealed that the second band thickness is much thinner than the ISe thickness observed in clinical OCT [115]. The second band thickness in the clinical OCT is consistent with the ISe thickness (16–20 µm) [66], but AO-OCT measurement demonstrated that the second band thickness is only about 4.7 µm [115]. Moreover, AO-OCT measurement of individual photoreceptors revealed that the second band peak is closer to the third band than the first band. Based on these observations, it was proposed that the second band in AO-OCT reflects the IS/OS junction instead of the ISe. A similar trend was also observed in the third band. The third band thickness was around 4.3–6.4 µm in AO-OCT, while the third band thickness in clinical OCT was around 14 to 19 µm [116]. Consequently, the AO-OCT studies suggested that the band thickness in clinical OCT may be overestimated. Yao et al. suggested that the AO-OCT can selectively enhance the sensitivity for imaging ballistic photons from the IS/OS junction, while partially rejecting the diffusive photons within the ISe region due to higher sectioning capability [66]. Thus, in clinical OCT, both IS/OS junction and ISe can nonexclusively contribute to the second band; and OS, OS tips, and RPE apical processes contribute to the third band. Also, different weighting factors, due to the different system resolution, aberration, effective pupil size, illumination, imaging orientation can simultaneously contribute to the signal detection in clinical OCT and AO-OCT. Therefore, it should be acknowledged that the contributing factors for individual band correlates are variable in different instruments, testing protocols, and eye conditions.

## 5.2 Future Perspectives

Growing evidence indicates that morphological examination may be limited to detect the early stage of retinal degenerative diseases, including but not limited to AMD [117], DR [118], and IRDs [119]. Given that timely management of the diseases is the key to preservation of vision [120; 121; 122; 123], functional assessment of retinal photoreceptors and neurovascular coupling has gained increasing importance. Ongoing development in OCT based ORG measurement is one of the most promising methodologies for screening people at risk. With unparallel depth-resolved capability, recent advances in OCT further provide multi-modalities, ultrahigh speed recording, single-cell resolution, and ultrawide field recording. Moreover, advanced ORG processing algorithms allows mapping of various functional activities over the morphological images. While the implementation of OCT-ORG in clinics is still at an early stage, recent studies demonstrated the feasibility of ORG measurement in human subjects [25; 31; 32; 36; 37; 54; 56; 68; 69; 100; 103]. To facilitate clinical transition, it would be necessary to refine experimental procedures and shorten the examination time including the light/dark adaptation to reduce the subject's burden. In addition, there is a broad range of existing testing methods for functional examination, and each has benefits and limitations. In this regard, OCT-ORG should be conducted alongside the existing tests as a multimodal evaluation, which can provide a better understanding of retinal physiology and corresponding IOS sources. Standardized imaging protocol and processing methods also need to be established as there are significant variations in imaging quality, appearance of the retina and following results due to different systems and processing algorithms. Above all, our understanding of functional activity and the corresponding IOS change is quite limited. Thus, there is an unmet need to seek direct evidence of biological processes to visual stimulus, which help translate distinct intrinsic signal sources at different retinal locations to target retinal disorders. Both *in vivo* and *ex vivo* study using mutant animal models would be useful. We anticipate that further development of the OCT system and ORG processing methods promises an objective measurement of neural and hemodynamic dysfunctions in the retina, allowing early detection and therapeutic assessment of AMD, DR, IRDs and other retinal diseases.

## Author Contributions

TK conceived the article, performed the literature search, drafted the manuscript, and prepared figures. GM and TS edited the manuscript and prepared figures. XY conceived the article, edited the manuscript, and supervised the study. All authors approved the submitted version.

## Funding

National Eye Institute: P30 EY001792, R01 EY030101, R01 EY023522, R01EY029673, R01 EY030842, R44 EY028786; Richard and Loan Hill endowment; unrestricted grant from Research to Prevent Blindness.

## Conflict of Interest

The authors declare that the research was conducted in the absence of any commercial or financial relationships that could be construed as a potential conflict of interest.

## References

[1] X. Li, J. Xie, L. Zhang, Y. Cui, G. Zhang, J. Wang, A. Zhang, X. Chen, T. Huang, and Q. Meng, Differential distribution of manifest lesions in diabetic retinopathy by fundus fluorescein angiography and fundus photography. BMC Ophthalmol 20 (2020) 471.

[2] D.E. Croft, J. van Hemert, C.C. Wykoff, D. Clifton, M. Verhoek, A. Fleming, and D.M. Brown, Precise montaging and metric quantification of retinal surface area from ultra-widefield fundus photography and fluorescein angiography. Ophthalmic Surg Lasers Imaging Retina 45 (2014) 312-7.

[3] J.H. Lee, S.S. Kim, and G.T. Kim, Microvascular findings in patients with systemic lupus erythematosus assessed by fundus photography with fluorescein angiography. Clin Exp Rheumatol 31 (2013) 871-6.

[4] J. Fischer, T. Otto, F. Delori, L. Pace, and G. Staurenghi, Scanning Laser Ophthalmoscopy (SLO). in: J.F. Bille, (Ed.), High Resolution Imaging in Microscopy and Ophthalmology: New Frontiers in Biomedical Optics, Cham (CH), 2019, pp. 35-57.

[5] A.M. Calvo-Maroto, J.J. Esteve-Taboada, A. Dominguez-Vicent, R.J. Perez-Cambrodi, and A. Cervino, Confocal scanning laser ophthalmoscopy versus modified conventional fundus camera for fundus autofluorescence. Expert Rev Med Devices 13 (2016) 965-978.

[6] E.A. Swanson, J.A. Izatt, M.R. Hee, D. Huang, C.P. Lin, J.S. Schuman, C.A. Puliafito, and J.G. Fujimoto, In vivo retinal imaging by optical coherence tomography. Opt Lett 18 (1993) 1864-6.

[7] R.A. Leitgeb, En face optical coherence tomography: a technology review [Invited]. Biomed. Opt. Express 10 (2019) 2177-2201.

[8] R.F. Spaide, J.G. Fujimoto, N.K. Waheed, S.R. Sadda, and G. Staurenghi, Optical coherence tomography angiography. Prog Retin Eye Res 64 (2018) 1-55.

[9] C.L. Chen, and R.K. Wang, Optical coherence tomography based angiography [Invited]. Biomed Opt Express 8 (2017) 1056-1082.

[10] S.S. Gao, Y. Jia, M. Zhang, J.P. Su, G. Liu, T.S. Hwang, S.T. Bailey, and D. Huang, Optical Coherence Tomography Angiography. Invest Ophthalmol Vis Sci 57 (2016) OCT27-36.

[11] X. Yao, M.N. Alam, D. Le, and D. Toslak, Quantitative optical coherence tomography angiography: A review. Exp Biol Med (Maywood) 245 (2020) 301-312.

[12] K. Bessho, T. Fujikado, T. Mihashi, T. Yamaguchi, N. Nakazawa, and Y. Tano, Photoreceptor images of normal eyes and of eyes with macular dystrophy obtained in vivo with an adaptive optics fundus camera. Jpn J Ophthalmol 52 (2008) 380-385.

[13] M.K. Soliman, M.A. Sadiq, A. Agarwal, S. Sarwar, M. Hassan, M. Hanout, F. Graf, R. High, D.V. Do, Q.D. Nguyen, and Y.J. Sepah, High-Resolution Imaging of Parafoveal Cones in Different Stages of Diabetic Retinopathy Using Adaptive Optics Fundus Camera. PLoS One 11 (2016) e0152788.

[14] B. Zhang, N. Li, J. Kang, Y. He, and X.M. Chen, Adaptive optics scanning laser ophthalmoscopy in fundus imaging, a review and update. Int J Ophthalmol 10 (2017) 1751-1758.

[15] A. Roorda, Applications of adaptive optics scanning laser ophthalmoscopy. Optom Vis Sci 87 (2010) 260-8.

[16] P. Godara, C. Siebe, J. Rha, M. Michaelides, and J. Carroll, Assessing the photoreceptor mosaic over drusen using adaptive optics and SD-OCT. Ophthalmic Surg Lasers Imaging 41 Suppl (2010) S104-8.

[17] B.J. King, K.A. Sapoznik, A.E. Elsner, T.J. Gast, J.A. Papay, C.A. Clark, and S.A. Burns, SD-OCT and Adaptive Optics Imaging of Outer Retinal Tubulation. Optom Vis Sci 94 (2017) 411-422.

[18] D.T. Miller, O.P. Kocaoglu, Q. Wang, and S. Lee, Adaptive optics and the eye (super resolution OCT). Eye (Lond) 25 (2011) 321-30.

[19] T.Y. Lai, W.M. Chan, R.Y. Lai, J.W. Ngai, H. Li, and D.S. Lam, The clinical applications of multifocal electroretinography: a systematic review. Surv Ophthalmol 52 (2007) 61-96.

[20] J.J. McAnany, O.S. Persidina, and J.C. Park, Clinical electroretinography in diabetic retinopathy: a review. Surv Ophthalmol (2021).

[21] D. Ts'o, J. Schallek, Y. Kwon, R. Kardon, M. Abramoff, and P. Soliz, Noninvasive functional imaging of the retina reveals outer retinal and hemodynamic intrinsic optical signal origins. Jpn J Ophthalmol 53 (2009) 334-44.

[22] X. Yao, and B. Wang, Intrinsic optical signal imaging of retinal physiology: a review. J Biomed Opt 20 (2015) 090901.

[23] G. Hanazono, K. Tsunoda, Y. Kazato, K. Tsubota, and M. Tanifuji, Evaluating neural activity of retinal ganglion cells by flash-evoked intrinsic signal imaging in macaque retina. Invest Ophthalmol Vis Sci 49 (2008) 4655-63.

[24] M. Begum, D.P. Joiner, and D.Y. Ts'o, Stimulus-Driven Retinal Intrinsic Signal Optical Imaging in Mouse Demonstrates a Dominant Rod-Driven Component. Invest Ophthalmol Vis Sci 61 (2020) 37.

[25] T. Son, T.H. Kim, G. Ma, H. Kim, and X. Yao, Functional intrinsic optical signal imaging for objective optoretinography of human photoreceptors. Exp Biol Med (Maywood) 246 (2021) 639-643.

[26] B. Wang, and X. Yao, In vivo intrinsic optical signal imaging of mouse retinas. Proc SPIE Int Soc Opt Eng 9693 (2016).

[27] Q.X. Zhang, Y. Zhang, R.W. Lu, Y.C. Li, S.J. Pittler, T.W. Kraft, and X.C. Yao, Comparative intrinsic optical signal imaging of wild-type and mutant mouse retinas. Opt Express 20 (2012) 7646-54.

[28] X.C. Yao, Intrinsic optical signal imaging of retinal activation. Jpn J Ophthalmol 53 (2009) 327-33.

[29] X.C. Yao, A. Yamauchi, B. Perry, and J.S. George, Rapid optical coherence tomography and recording functional scattering changes from activated frog retina. Applied Optics 44 (2005) 2019-2023.

[30] K. Bizheva, R. Pflug, B. Hermann, B. Povazay, H. Sattmann, P. Qiu, E. Anger, H. Reitsamer, S. Popov, J.R. Taylor, A. Unterhuber, P. Ahnelt, and W. Drexler, Optophysiology: depth-resolved probing of retinal physiology with functional ultrahigh-resolution optical coherence tomography. Proc Natl Acad Sci U S A 103 (2006) 5066-71.

[31] M. Azimipour, D. Valente, K.V. Vienola, J.S. Werner, R.J. Zawadzki, and R.S. Jonnal, Optoretinogram: optical measurement of human cone and rod photoreceptor responses to light. Opt Lett 45 (2020) 4658-4661.

[32] V.P. Pandiyan, A. Maloney-Bertelli, J.A. Kuchenbecker, K.C. Boyle, T. Ling, Z.C. Chen, B.H. Park, A. Roorda, D. Palanker, and R. Sabesan, The optoretinogram reveals the primary steps of phototransduction in the living human eye. Sci Adv 6 (2020).

[33] L. Zhang, R. Dong, R.J. Zawadzki, and P. Zhang, Volumetric data analysis enabled spatially resolved optoretinogram to measure the functional signals in the living retina. J Biophotonics (2021) e202100252.

[34] R.S. Jonnal, Toward a clinical optoretinogram: a review of noninvasive, optical tests of retinal neural function. Ann Transl Med 9 (2021) 1270.

[35] P. Zhang, B. Shibata, G. Peinado, R.J. Zawadzki, P. FitzGerald, and E.N. Pugh, Jr., Measurement of Diurnal Variation in Rod Outer Segment Length In Vivo in Mice With the OCT Optoretinogram. Invest Ophthalmol Vis Sci 61 (2020) 9.

[36] A. Lassoued, F. Zhang, K. Kurokawa, Y. Liu, M.T. Bernucci, J.A. Crowell, and D.T. Miller, Cone photoreceptor dysfunction in retinitis pigmentosa revealed by optoretinography. Proc Natl Acad Sci U S A 118 (2021).

[37] G. Ma, T. Son, T.H. Kim, and X. Yao, Functional optoretinography: concurrent OCT monitoring of intrinsic signal amplitude and phase dynamics in human photoreceptors. Biomed Opt Express 12 (2021) 2661-2669.

[38] R.F. Cooper, D.H. Brainard, and J.I.W. Morgan, Optoretinography of individual human cone photoreceptors. Opt Express 28 (2020) 39326-39339.

[39] T.H. Kim, B. Wang, Y. Lu, T. Son, and X. Yao, Functional optical coherence tomography enables in vivo optoretinography of photoreceptor dysfunction due to retinal degeneration. Biomed Opt Express 11 (2020) 5306-5320.

[40] V.P. Pandiyan, X. Jiang, A. Maloney-Bertelli, J.A. Kuchenbecker, U. Sharma, and R. Sabesan, High-speed adaptive optics line-scan OCT for cellular-resolution optoretinography. Biomed Opt Express 11 (2020) 5274-5296.

[41] X. Yao, T. Son, T.H. Kim, and Y. Lu, Functional optical coherence tomography of retinal photoreceptors. Exp Biol Med (Maywood) 243 (2018) 1256-1264.

[42] Q. Zhang, R. Lu, B. Wang, J.D. Messinger, C.A. Curcio, and X. Yao, Functional optical coherence tomography enables in vivo physiological assessment of retinal rod and cone photoreceptors. Sci Rep 5 (2015) 9595.

[43] B. Wang, R. Lu, Q. Zhang, Y. Jiang, and X. Yao, En face optical coherence tomography of transient light response at photoreceptor outer segments in living frog eyecup. Opt Lett 38 (2013) 4526-9.

[44] A.A. Moayed, S. Hariri, V. Choh, and K. Bizheva, In vivo imaging of intrinsic optical signals in chicken retina with functional optical coherence tomography. Opt Lett 36 (2011) 4575-7.

[45] X. Yao, and T.H. Kim, Fast intrinsic optical signal correlates with activation phase of phototransduction in retinal photoreceptors. Exp Biol Med (Maywood) 245 (2020) 1087-1095.

[46] V.J. Srinivasan, M. Wojtkowski, J.G. Fujimoto, and J.S. Duker, In vivo measurement of retinal physiology with high-speed ultrahigh-resolution optical coherence tomography. Opt Lett 31 (2006) 2308-10.

[47] B. Wang, Q. Zhang, R. Lu, Y. Zhi, and X. Yao, Functional optical coherence tomography reveals transient phototropic change of photoreceptor outer segments. Opt Lett 39 (2014) 6923-6.

[48] A. Akhlagh Moayed, S. Hariri, V. Choh, and K. Bizheva, Correlation of visually evoked intrinsic optical signals and electroretinograms recorded from chicken retina with a combined functional optical coherence tomography and electroretinography system. J Biomed Opt 17 (2012) 016011.

[49] Q.X. Zhang, R.W. Lu, J.D. Messinger, C.A. Curcio, V. Guarcello, and X.C. Yao, In vivo Optical Coherence Tomography of Light-Driven Melanosome Translocation in Retinal Pigment Epithelium. Sci Rep 3 (2013) 2644.

[50] T. Son, B. Wang, D. Thapa, Y. Lu, Y. Chen, D. Cao, and X. Yao, Optical coherence tomography angiography of stimulus evoked hemodynamic responses in individual retinal layers. Biomed Opt Express 7 (2016) 3151-62.

[51] T. Son, B. Wang, Y. Lu, Y. Chen, D. Cao, and X. Yao, Concurrent OCT imaging of stimulus evoked retinal neural activation and hemodynamic responses. Proc SPIE Int Soc Opt Eng 10045 (2017).

[52] W. Suzuki, K. Tsunoda, G. Hanazono, and M. Tanifuji, Stimulus-induced changes of reflectivity detected by optical coherence tomography in macaque retina. Invest Ophthalmol Vis Sci 54 (2013) 6345-54.

[53] V.J. Srinivasan, Y. Chen, J.S. Duker, and J.G. Fujimoto, In vivo functional imaging of intrinsic scattering changes in the human retina with high-speed ultrahigh resolution OCT. Opt Express 17 (2009) 3861-77.

[54] A. Messner, V. Aranha dos Santos, H. Stegmann, S. Puchner, D. Schmidl, R. Leitgeb, L. Schmetterer, and R.M. Werkmeister, Quantification of intrinsic optical signals in the outer human retina using optical coherence tomography. Annals of the New York Academy of Sciences n/a (2021).

[55] D. Hillmann, C. Pfäffle, H. Spahr, H. Sudkamp, G. Franke, and G. Hüttmann, In Vivo FF-SS-OCT Optical Imaging of Physiological Responses to Photostimulation of Human Photoreceptor Cells. in: J.F. Bille, (Ed.), High Resolution Imaging in Microscopy and Ophthalmology: New Frontiers in Biomedical Optics, Springer International Publishing, Cham, 2019, pp. 181-194.

[56] D. Hillmann, H. Spahr, C. Pfaffle, H. Sudkamp, G. Franke, and G. Huttmann, In vivo optical imaging of physiological responses to photostimulation in human photoreceptors. Proc Natl Acad Sci U S A 113 (2016) 13138-13143.

[57] R.S. Jonnal, O.P. Kocaoglu, R.J. Zawadzki, Z. Liu, D.T. Miller, and J.S. Werner, A Review of Adaptive Optics Optical Coherence Tomography: Technical Advances, Scientific Applications, and the Future. Investigative Ophthalmology & Visual Science 57 (2016) OCT51-OCT68.

[58] K. Tae-Hoon, W. Benquan, L. Yiming, S. Taeyoon, and Y. Xincheng, Intrinsic signal optoretinography of rod photoreceptor dysfunction due to retinal degeneration, Proc.SPIE, 2021.

[59] G. Ma, T. Son, T.H. Kim, and X. Yao, In vivo optoretinography of phototransduction activation and energy metabolism in retinal photoreceptors. J Biophotonics 14 (2021) e202000462.

[60] B. Wang, Y. Lu, and X. Yao, In vivo optical coherence tomography of stimulus-evoked intrinsic optical signals in mouse retinas. J Biomed Opt 21 (2016) 96010.

[61] T. Son, M. Alam, D. Toslak, B. Wang, Y. Lu, and X. Yao, Functional optical coherence tomography of neurovascular coupling interactions in the retina. J Biophotonics 11 (2018) e201800089.

[62] T.-H. Kim, B. Wang, Y. Lu, T. Son, and X. Yao, Retinal intrinsic optical signal imaging of wild-type and rd10 mice. Investigative Ophthalmology & Visual Science 61 (2020) 198-198.

[63] T.-H. Kim, J. Ding, and X. Yao, Intrinsic signal optoretinography of dark adaptation kinetics. arXiv e-prints (2021) arXiv:2112.07838.

[64] P. Zhang, R.J. Zawadzki, M. Goswami, P.T. Nguyen, V. Yarov-Yarovoy, M.E. Burns, and E.N. Pugh, In vivo optophysiology reveals that G-protein activation triggers osmotic swelling and increased light scattering of rod photoreceptors. Proceedings of the National Academy of Sciences 114 (2017) E2937.

[65] Y. Li, Y. Zhang, S. Chen, G. Vernon, W.T. Wong, and H. Qian, Light-Dependent OCT Structure Changes in Photoreceptor Degenerative rd 10 Mouse Retina. Investigative Ophthalmology & Visual Science 59 (2018) 1084-1094.

[66] X. Yao, T. Son, T.-H. Kim, and D. Le, Interpretation of anatomic correlates of outer retinal bands in optical coherence tomography. Experimental Biology and Medicine 246 (2021) 2140-2150.

[67] A. Roorda, and D.R. Williams, Optical fiber properties of individual human cones. Journal of Vision 2 (2002) 4-4.

[68] F. Zhang, K. Kurokawa, A. Lassoued, J.A. Crowell, and D.T. Miller, Cone photoreceptor classification in the living human eye from photostimulation-induced phase dynamics. Proceedings of the National Academy of Sciences 116 (2019) 7951.

[69] C. Pfäffle, H. Spahr, L. Kutzner, S. Burhan, F. Hilge, Y. Miura, G. Hüttmann, and D. Hillmann, Simultaneous functional imaging of neuronal and photoreceptor layers in living human retina. Opt. Lett. 44 (2019) 5671-5674.

[70] A.V. Tschulakow, T. Oltrup, T. Bende, S. Schmelzle, and U. Schraermeyer, The anatomy of the foveola reinvestigated. PeerJ 6 (2018) e4482.

[71] C.E. Riva, E. Logean, and B. Falsini, Visually evoked hemodynamical response and assessment of neurovascular coupling in the optic nerve and retina. Progress in Retinal and Eye Research 24 (2005) 183-215.

[72] L.S. Lim, L.H. Ling, P.G. Ong, W. Foulds, E.S. Tai, and T.Y. Wong, Dynamic Responses in Retinal Vessel Caliber With Flicker Light Stimulation and Risk of Diabetic Retinopathy and Its Progression. Investigative Ophthalmology & Visual Science 58 (2017) 2449-2455.

[73] A. Mandecka, J. Dawczynski, M. Blum, N. Müller, C. Kloos, G. Wolf, W. Vilser, H. Hoyer, and U.A. Müller, Influence of Flickering Light on the Retinal Vessels in Diabetic Patients. Diabetes Care 30 (2007) 3048-3052.

[74] T. Zhang, A.M. Kho, and V.J. Srinivasan, In vivo Morphometry of Inner Plexiform Layer (IPL) Stratification in the Human Retina With Visible Light Optical Coherence Tomography. Frontiers in Cellular Neuroscience 15 (2021) 147.

[75] Z. Ghassabi, R.V. Kuranov, J.S. Schuman, R. Zambrano, M. Wu, M. Liu, B. Tayebi, Y. Wang, I. Rubinoff, X. Liu, G. Wollstein, H.F. Zhang, and H. Ishikawa, In Vivo Sublayer Analysis of Human Retinal Inner Plexiform Layer Obtained by Visible-Light Optical Coherence Tomography. Investigative Ophthalmology & Visual Science 63 (2022) 18-18.

[76] Y.-B. Zhao, and X.-C. Yao, Intrinsic optical imaging of stimulus-modulated physiological responses in amphibian retina. Opt. Lett. 33 (2008) 342-344.

[77] Y. Xincheng, and S.G. John, Near-infrared imaging of fast intrinsic optical responses in visible light-activated amphibian retina. Journal of Biomedical Optics 11 (2006) 1-8.

[78] L. Yiming, L. Changgeng, and Y. Xincheng, In vivo super-resolution imaging of transient retinal phototropism evoked by oblique light stimulation. Journal of Biomedical Optics 23 (2018) 1-4.

[79] L. Rongwen, M.L. Alexander, Z. Qiuxiang, J.P. Steven, and Y. Xincheng, Dynamic near-infrared imaging reveals transient phototropic change in retinal rod photoreceptors. Journal of Biomedical Optics 18 (2013) 1-7.

[80] Y. Lu, J. Benedetti, and X. Yao, Light-Induced Length Shrinkage of Rod Photoreceptor Outer Segments. Translational Vision Science & Technology 7 (2018) 29-29.

[81] Y. Lu, B. Wang, D.R. Pepperberg, and X. Yao, Stimulus-evoked outer segment changes occur before the hyperpolarization of retinal photoreceptors. Biomed Opt Express 8 (2017) 38-47.

[82] Y. Lu, T.-H. Kim, and X. Yao, Comparative study of wild-type and rd10 mice reveals transient intrinsic optical signal response before phosphodiesterase activation in retinal photoreceptors. Experimental Biology and Medicine 245 (2019) 360-367.

[83] K.C. Boyle, Z.C. Chen, T. Ling, V.P. Pandiyan, J. Kuchenbecker, R. Sabesan, and D. Palanker, Mechanisms of Light-Induced Deformations in Photoreceptors. Biophysical Journal 119 (2020) 1481-1488.

[84] K.T. Brown, and M. Murakami, A New Receptor Potential of the Monkey Retina with no Detectable Latency. Nature 201 (1964) 626-628.

[85] S. Hestrin, and J.I. Korenbrot, Activation kinetics of retinal cones and rods: response to intense flashes of light. The Journal of Neuroscience 10 (1990) 1967.

[86] C.D. Lu, B. Lee, J. Schottenhamml, A. Maier, E.N. Pugh, Jr., and J.G. Fujimoto, Photoreceptor Layer Thickness Changes During Dark Adaptation Observed With Ultrahigh-Resolution Optical Coherence Tomography. Investigative Ophthalmology & Visual Science 58 (2017) 4632-4643.

[87] M. Chabre, X-ray diffraction studies of retinal rods. I. Structure of the disc membrane, effect of illumination. Biochimica et Biophysica Acta (BBA) - Biomembranes 382 (1975) 322-335.

[88] N. Yagi, Structural changes in rod outer segments of frog and mouse after illumination. Experimental Eye Research 116 (2013) 395-401.

[89] U. Bocchero, F. Falleroni, S. Mortal, Y. Li, D. Cojoc, T. Lamb, and V. Torre, Mechanosensitivity is an essential component of phototransduction in vertebrate rods. PLOS Biology 18 (2020) e3000750.

[90] J.N. Pearring, R.Y. Salinas, S.A. Baker, and V.Y. Arshavsky, Protein sorting, targeting and trafficking in photoreceptor cells. Progress in Retinal and Eye Research 36 (2013) 24-51.

[91] A. Morshedian, and G.L. Fain, The evolution of rod photoreceptors. Philosophical Transactions of the Royal Society B: Biological Sciences 372 (2017) 20160074.

[92] Z. Xiaohui, T. Damber, W. Benquan, L. Yiming, G. Shaoyan, and Y. Xincheng, Stimulus-evoked outer segment changes in rod photoreceptors. Journal of Biomedical Optics 21 (2016) 1-8.

[93] J. Stone, D. van Driel, K. Valter, S. Rees, and J. Provis, The locations of mitochondria in mammalian photoreceptors: Relation to retinal vasculature. Brain Research 1189 (2008) 58-69.

[94] K.E. Lee, H. Heitkotter, and J. Carroll, Challenges Associated With Ellipsoid Zone Intensity Measurements Using Optical Coherence Tomography. Transl Vis Sci Technol 10 (2021) 27.

[95] B. Westermann, Mitochondrial fusion and fission in cell life and death. Nature Reviews Molecular Cell Biology 11 (2010) 872-884.

[96] N.-A. Pham, T. Richardson, J. Cameron, B. Chue, and B.H. Robinson, Altered Mitochondrial Structure and Motion Dynamics in Living Cells with Energy Metabolism Defects Revealed by Real Time Microscope Imaging. Microscopy and Microanalysis 10 (2004) 247-260.

[97] B.A. Berkowitz, and H. Qian, OCT imaging of rod mitochondrial respiration in vivo. Exp Biol Med (Maywood) 246 (2021) 2151-2158.

[98] Y. Li, R.N. Fariss, J.W. Qian, E.D. Cohen, and H. Qian, Light-Induced Thickening of Photoreceptor Outer Segment Layer Detected by Ultra-High Resolution OCT Imaging. Investigative Ophthalmology & Visual Science 57 (2016) OCT105-OCT111.

[99] B.A. Berkowitz, R.H. Podolsky, H. Qian, Y. Li, K. Jiang, J. Nellissery, A. Swaroop, and R. Roberts, Mitochondrial Respiration in Outer Retina Contributes to Light-Evoked Increase in Hydration In Vivo. Investigative Ophthalmology & Visual Science 59 (2018) 5957-5964.

[100] S. Gao, Y. Li, D. Bissig, E.D. Cohen, R.H. Podolsky, K.L. Childers, G. Vernon, S. Chen, B.A. Berkowitz, and H. Qian, Functional regulation of an outer retina hyporeflective band on optical coherence tomography images. Scientific Reports 11 (2021) 10260.

[101] J. Adijanto, T. Banzon, S. Jalickee, N.S. Wang, and S.S. Miller, CO2-induced ion and fluid transport in human retinal pigment epithelium. Journal of General Physiology 133 (2009) 603-622.

[102] N.D. Wangsa-Wirawan, and R.A. Linsenmeier, Retinal Oxygen: Fundamental and Clinical Aspects. Archives of Ophthalmology 121 (2003) 547-557.

[103] A. Messner, R.M. Werkmeister, G. Seidel, H. Stegmann, L. Schmetterer, and V. Aranha dos Santos, Light-induced changes of the subretinal space of the temporal retina observed via optical coherence tomography. Scientific Reports 9 (2019) 13632.

[104] M. Azimipour, J.V. Migacz, R.J. Zawadzki, J.S. Werner, and R.S. Jonnal, Functional retinal imaging using adaptive optics swept-source OCT at 1.6 MHz. Optica 6 (2019) 300-303.

[105] B.J. Lujan, A. Roorda, R.W. Knighton, and J. Carroll, Revealing Henle's Fiber Layer Using Spectral Domain Optical Coherence Tomography. Investigative Ophthalmology & Visual Science 52 (2011) 1486-1492.

[106] W. Gao, B. Cense, Y. Zhang, R.S. Jonnal, and D.T. Miller, Measuring retinal contributions to the optical Stiles-Crawford effect with optical coherence tomography. Opt. Express 16 (2008) 6486-6501.

[107] K.M. Litts, Y. Zhang, K.B. Freund, and C.A. Curcio, OPTICAL COHERENCE TOMOGRAPHY AND HISTOLOGY OF AGE-RELATED MACULAR DEGENERATION SUPPORT MITOCHONDRIA AS REFLECTIVITY SOURCES. RETINA 38 (2018).

[108] S. Thiele, B. Isselmann, M. Pfau, F.G. Holz, S. Schmitz-Valckenberg, Z. Wu, R.H. Guymer, and C.D. Luu, Validation of an Automated Quantification of Relative Ellipsoid Zone Reflectivity on Spectral Domain-Optical Coherence Tomography Images. Translational Vision Science & Technology 9 (2020) 17-17.

[109] Z. Hu, A. Hariri, X. Wu, and S.R. Sadda, Comparison of retinal layer profiles between spectral-domain optical coherence tomography devices. Investigative Ophthalmology & Visual Science 55 (2014) 4785-4785.

[110] M. Ratheesh Kumar, Z. Pengfei, J. Myeong Jin, K.K.M. Suman, J. Yifan, N.P. Edward, and J.Z. Robert, Directional optical coherence tomography reveals melanin concentration-dependent scattering properties of retinal pigment epithelium. Journal of Biomedical Optics 24 (2019) 1-10.

[111] S.M. Bloom, and I.P. Singal, REVISED CLASSIFICATION OF THE OPTICAL COHERENCE TOMOGRAPHY OUTER RETINAL BANDS BASED ON CENTRAL SEROUS CHORIORETINOPATHY ANALYSIS. RETINA 41 (2021).

[112] N. Cuenca, I. Ortuño-Lizarán, and I. Pinilla, Cellular Characterization of OCT and Outer Retinal Bands Using Specific Immunohistochemistry Markers and Clinical Implications. Ophthalmology 125 (2018) 407-422.

[113] R.F. Spaide, and C.A. Curcio, ANATOMICAL CORRELATES TO THE BANDS SEEN IN THE OUTER RETINA BY OPTICAL COHERENCE TOMOGRAPHY: Literature Review and Model. RETINA 31 (2011).

[114] G. Staurenghi, S. Sadda, U. Chakravarthy, and R.F. Spaide, Proposed Lexicon for Anatomic Landmarks in Normal Posterior Segment Spectral-Domain Optical Coherence Tomography: The IN•OCT Consensus. Ophthalmology 121 (2014) 1572-1578.

[115] R.S. Jonnal, O.P. Kocaoglu, R.J. Zawadzki, S.-H. Lee, J.S. Werner, and D.T. Miller, The Cellular Origins of the Outer Retinal Bands in Optical Coherence Tomography Images. Investigative Ophthalmology & Visual Science 55 (2014) 7904-7918.

[116] R.S. Jonnal, I. Gorczynska, J.V. Migacz, M. Azimipour, R.J. Zawadzki, and J.S. Werner, The Properties of Outer Retinal Band Three Investigated With Adaptive-Optics Optical Coherence Tomography. Investigative Ophthalmology & Visual Science 58 (2017) 4559-4568.

[117] W.M. Haddad, A. Seres, G. Coscas, and G. Soubrane, Presentation delay in patients affected with exudative age-related macular degeneration. Graefe's Archive for Clinical and Experimental Ophthalmology 240 (2002) 31-34.

[118] S.C. Lee, E.T. Lee, R.M. Kingsley, Y. Wang, D. Russell, R. Klein, and A. Warn, Comparison of Diagnosis of Early Retinal Lesions of Diabetic Retinopathy Between a Computer System and Human Experts. Archives of Ophthalmology 119 (2001) 509-515.

[119] A. Suppiej, S. Marino, M.E. Reffo, V. Maritan, G. Vitaliti, J. Mailo, and R. Falsaperla, Early onset retinal dystrophies: clinical clues to diagnosis for pediatricians. Italian Journal of Pediatrics 45 (2019) 168.

[120] G. The Vegf Inhibition Study In Ocular Neovascularization Clinical Trial, ENHANCED EFFICACY ASSOCIATED WITH EARLY TREATMENT OF NEOVASCULAR AGE-RELATED MACULAR DEGENERATION WITH PEGAPTANIB SODIUM: An Exploratory Analysis. RETINA 25 (2005).

[121] A. Loewenstein, THE SIGNIFICANCE OF EARLY DETECTION OF AGE-RELATED MACULAR DEGENERATION: Richard & Hinda Rosenthal Foundation Lecture, The Macula Society 29th Annual Meeting. RETINA 27 (2007).

[122] K. Oh, H.M. Kang, D. Leem, H. Lee, K.Y. Seo, and S. Yoon, Early detection of diabetic retinopathy based on deep learning and ultra-wide-field fundus images. Scientific Reports 11 (2021) 1897.

[123] B.P. Hafler, CLINICAL PROGRESS IN INHERITED RETINAL DEGENERATIONS: GENE THERAPY CLINICAL TRIALS AND ADVANCES IN GENETIC SEQUENCING. RETINA 37 (2017).